\documentclass[]{aastex701}
\usepackage{color}

\newcommand{\spherex}{SPHEREx}
\newcommand{\launchMonth}{March~2025}
\newcommand{\ssdc}{SSDC}
\newcommand{\ssdcFull}{\spherex\ Science Data Center}
\newcommand{\wise}{WISE}
\newcommand{\wxscFull}{\wise\ Extended Source Catalog}
\newcommand{\wxsc}{WXSC}
\newcommand{\twomass}{2MASS}
\newcommand{\healpix}{HEALPix}
\newcommand{\irsa}{IRSA}
\newcommand{\irsaFull}{NASA/IPAC Infrared Science Archive}
\newcommand{\HI}{H\,{\sc i}}
\newcommand{\HII}{H\,{\sc ii}}
\newcommand{\HeI}{He\,{\sc i}}
\newcommand{\SIII}{S\,{\sc iii}}
\newcommand{\ArIII}{Ar\,{\sc iii}}
\newcommand{\Hmol}{H$_2$}
\newcommand{\CO}{CO}
\newcommand{\COtwo}{CO$_2$}
\newcommand{\water}{H$_2$O}
\newcommand{\um}{\mu\mathrm{m}}
\newcommand{\km}{\mathrm{km}}
\newcommand{\figenv}[1]{\begin{figure}\begin{center}#1\end{center}\end{figure}}
\newcommand{\figref}[1]{Fig.~\ref{#1}}
\newcommand{\secref}[1]{Sec.~\ref{#1}}
\newcommand{\eqref}[1]{Eq.~\ref{#1}}
\newcommand{\secs}[1]{Secs.~#1}
\newcommand{\figs}[1]{Figs.~#1}
\newcommand{\fnref}[1]{footnote~\ref{#1}}
\newcommand{\nside}{N_\mathrm{side}}
\newcommand{\wlHe}{1.083}
\newcommand{\order}[1]{\sim #1}
\newcommand{\mAB}{m_\mathrm{AB}}
\newcommand{\s}{\mathrm{s}}
\newcommand{\wlBra}{4.052} 
\newcommand{\Bra}{Brackett-$\alpha$}
\newcommand{\wlPfb}{4.654} 
\newcommand{\Pfb}{Pfund-$\beta$}
\newcommand{\wlPaa}{1.876} 
\newcommand{\Paa}{Paschen-$\alpha$}
\newcommand{\Pab}{Paschen-$\beta$}
\newcommand{\wlPAH}{3.3} 
\newcommand{\wlArIII}{0.775}
\newcommand{\wlSIII}{0.953}
\newcommand{\wlSIIIdim}{0.907}

\newcommand{\atlas}{3I/ATLAS}
\newcommand{\skysim}{Sky Simulator}

\newcommand{\st}{Science Team}
\newcommand{\MJysr}{\mathrm{MJy}/\mathrm{sr}}
\newcommand{\planck}{\emph{Planck}}
\newcommand{\Wone}{W1}
\newcommand{\wlWone}{3.4} 
\newcommand{\Wthree}{W3}
\newcommand{\wlWthree}{12} 
\newcommand{\milkyway}{Milky Way Galaxy}
\newcommand{\hiFourPi}{HI4PI}
\newcommand{\NHI}{N_{\text{\HI}}}
\newcommand{\cmInvTwo}{\mathrm{cm}^{-2}}
\newcommand{\NHIlinear}{4 \times 10^{20}~\cmInvTwo}
\newcommand{\fwhm}{\mathrm{FWHM}}
\newcommand{\neff}{N_\mathrm{eff}}
\renewcommand{\vec}[1]{\mathbf{#1}}
\newcommand{\wlZodiMin}{3.5} 

\newcommand{\nsidehits}{1024} 
\newcommand{\resolnsidehits}{3\farcm4} 
\newcommand{\fullskydate}{December~3, 2025}
\newcommand{\heart}{Heart Nebula}
\newcommand{\heartcat}{Sh2-190} 
\newcommand{\heartsoul}{Heart and Soul Nebula}
\newcommand{\soulcat}{Sh2-199}
\newcommand{\pinwheel}{Pinwheel Galaxy}
\newcommand{\pinwheelcat}{M101}
\newcommand{\wlCoadd}{3.913} 
\newcommand{\dimsCoadd}{7\degr \times 4 \degr} 
\newcommand{\pixsizeCoadd}{9\arcsec} 
\newcommand{\wlDiff}{4.176} 
\newcommand{\pixsizeDeep}{3\arcsec} 
\newcommand{\dimsReproj}{2\farcm5 \times 2\farcm5} 
\newcommand{\daysReproj}{223} 
\newcommand{\dimsDeep}{8\degr \times 8\degr} 
\newcommand{\dimsDeepZoom}{0\fdg25 \times 0\fdg25} 
\newcommand{\wlDeep}{1.940} 
\newcommand{\wlSim}{0.803} 
\newcommand{\catseyeCat}{NGC~6543}
\newcommand{\catseye}{Cat's Eye Nebula}
\newcommand{\diamCatsEye}{5\arcmin} 
\newcommand{\rExtHalo}{$2$-$3\arcmin$} 
\newcommand{\nChsCatsEye}{100} 
\newcommand{\dimsCatsEye}{7\arcmin \times 7\arcmin} 
\newcommand{\wlHmolEx}{4.6947} 
\newcommand{\HmolTransEx}{0-0~S(9)} 
\newcommand{\tri}{Triangulum Galaxy}
\newcommand{\tricat}{M33}
\newcommand{\wlMinContsub}{3.124} 
\newcommand{\wlMaxContsub}{3.603} 
\newcommand{\helix}{Helix Nebula}
\newcommand{\helixcat}{NGC~7293}
\newcommand{\nsubchsInterp}{102} 
\newcommand{\wlChInterp}{4.6937} 
\newcommand{\orion}{Orion Nebula}
\newcommand{\orioncat}{M42}
\newcommand{\bubble}{Bubble Nebula}
\newcommand{\bubblecat}{NGC~7635}
\newcommand{\nChsFullSky}{96} 
\newcommand{\fwhmFullSky}{20\arcmin} 
\newcommand{\nsideFullSky}{512} 
\newcommand{\resolFullSky}{6\farcm9} 
\newcommand{\wlDeproj}{0.763} 
\newcommand{\latDeproj}{45\degr} 
\newcommand{\NHImask}{10^{21}~\cmInvTwo} 
\newcommand{\fskyHI}{71.8\%} 
\newcommand{\wlDust}{100} 
\newcommand{\nlonBins}{10} 
\newcommand{\nsideSmooth}{2048} 
\newcommand{\resolSmooth}{1\farcm7} 
\newcommand{\fwhmSmooth}{4\arcmin} 

\newcommand{\pixsize}{6\farcs15}
\newcommand{\wlmin}{0.75}
\newcommand{\wlmax}{5.0}
\newcommand{\Rmin}{35}
\newcommand{\Rmax}{130}
\newcommand{\nchs}{102}
\newcommand{\exptime}{116}
\newcommand{\ndetside}{2040}
\newcommand{\Rsmall}{40} 
\newcommand{\Rlarge}{120} 
\newcommand{\fov}{3\fdg5} 
\newcommand{\fovfp}{11\fdg5} 
\newcommand{\altitude}{650} 
\newcommand{\elatSDF}{-82\degr} 
\newcommand{\elonSDF}{44\degr} 
\newcommand{\nsubchs}{17} 
\newcommand{\nsubchsDouble}{34} 

\begin{document}
\title{Spectral Map Making with \spherex}

\author[0000-0002-7471-719X]{Ari~J.~Cukierman}%
\affiliation{Department of Physics, California Institute of Technology, 1200 E. California Boulevard, Pasadena, CA 91125, USA}%
\email[show]{ajcukier@caltech.edu}%
\author[0009-0000-3415-2203]{Shuang-Shuang~Chen}%
\affiliation{Department of Physics, California Institute of Technology, 1200 E. California Boulevard, Pasadena, CA 91125, USA}%
\email{schen6@caltech.edu}%
\author[0000-0002-3470-2954]{Jae~Hwan~Kang}%
\affiliation{Department of Physics, California Institute of Technology, 1200 E. California Boulevard, Pasadena, CA 91125, USA}%
\email{jkang7@caltech.edu}%
\author[0000-0003-3393-2819]{Mary~H.~Minasyan}%
\affiliation{Department of Physics, California Institute of Technology, 1200 E. California Boulevard, Pasadena, CA 91125, USA}%
\email{minasyan@caltech.edu}%
\author[0009-0002-0149-9328]{Giulia~Murgia}%
\affiliation{Department of Physics, California Institute of Technology, 1200 E. California Boulevard, Pasadena, CA 91125, USA}%
\email{gmurgia@caltech.edu}%
\author[0000-0002-5710-5212]{James~J.~Bock}%
\affiliation{Department of Physics, California Institute of Technology, 1200 E. California Boulevard, Pasadena, CA 91125, USA}%
\affiliation{Jet Propulsion Laboratory, California Institute of Technology, 4800 Oak Grove Drive, Pasadena, CA 91109, USA}%
\email{jjb@astro.caltech.edu}%
\author[0000-0001-5929-4187]{Tzu-Ching~Chang}%
\affiliation{Jet Propulsion Laboratory, California Institute of Technology, 4800 Oak Grove Drive, Pasadena, CA 91109, USA}%
\affiliation{Department of Physics, California Institute of Technology, 1200 E. California Boulevard, Pasadena, CA 91125, USA}%
\email{tzu@caltech.edu}%
\author[0000-0001-6320-261X]{Yi-Kuan~Chiang}%
\affiliation{Academia Sinica Institute of Astronomy and Astrophysics (ASIAA), No. 1, Section 4, Roosevelt Road, Taipei 10617, Taiwan}%
\email{ykchiang@asiaa.sinica.edu.tw}%
\author[0000-0002-4650-8518]{Brendan~P.~Crill}%
\affiliation{Jet Propulsion Laboratory, California Institute of Technology, 4800 Oak Grove Drive, Pasadena, CA 91109, USA}%
\affiliation{Department of Physics, California Institute of Technology, 1200 E. California Boulevard, Pasadena, CA 91125, USA}%
\email{bcrill@jpl.nasa.gov}%
\author[0000-0001-7432-2932]{Olivier~Dor\'{e}}%
\affiliation{Jet Propulsion Laboratory, California Institute of Technology, 4800 Oak Grove Drive, Pasadena, CA 91109, USA}%
\affiliation{Department of Physics, California Institute of Technology, 1200 E. California Boulevard, Pasadena, CA 91125, USA}%
\email{olivier.dore@caltech.edu }%
\author[0009-0002-0098-6183]{C.~Darren~Dowell}%
\affiliation{Jet Propulsion Laboratory, California Institute of Technology, 4800 Oak Grove Drive, Pasadena, CA 91109, USA}%
\affiliation{Department of Physics, California Institute of Technology, 1200 E. California Boulevard, Pasadena, CA 91125, USA}%
\email{charles.d.dowell@jpl.nasa.gov}%
\author[0000-0002-9382-9832]{Andreas~L.~Faisst}%
\affiliation{IPAC, California Institute of Technology, MC 100-22, 1200 E California Blvd, Pasadena, CA 91125, USA}%
\email{afaisst@caltech.edu}%
\author[0000-0002-5599-4650]{Joseph~L.~Hora}%
\affiliation{Center for Astrophysics $|$ Harvard \& Smithsonian, Optical and Infrared Astronomy Division, Cambridge, MA 01238, USA}%
\email{jhora@cfa.harvard.edu}%
\author[0000-0001-5812-1903]{Howard~Hui}%
\affiliation{Department of Physics, California Institute of Technology, 1200 E. California Boulevard, Pasadena, CA 91125, USA}%
\affiliation{Jet Propulsion Laboratory, California Institute of Technology, 4800 Oak Grove Drive, Pasadena, CA 91109, USA}%
\email{hhui@caltech.edu}%
\author[0000-0002-5016-050X]{Miju~Kang}%
\affiliation{Korea Astronomy and Space Science Institute (KASI), 776 Daedeok-daero, Yuseong-gu, Daejeon 34055, Republic of Korea}%
\email{mjkang@kasi.re.kr}%
\author[0009-0003-8869-3365]{Phil~M.~Korngut}%
\affiliation{Department of Physics, California Institute of Technology, 1200 E. California Boulevard, Pasadena, CA 91125, USA}%
\email{pkorngut@caltech.edu}%
\author[0000-0002-3808-7143]{Ho-Gyu~Lee}%
\affiliation{Korea Astronomy and Space Science Institute (KASI), 776 Daedeok-daero, Yuseong-gu, Daejeon 34055, Republic of Korea}%
\email{hglee@kasi.re.kr}%
\author[0000-0003-1954-5046]{Bomee~Lee}%
\affiliation{Korea Astronomy and Space Science Institute (KASI), 776 Daedeok-daero, Yuseong-gu, Daejeon 34055, Republic of Korea}%
\affiliation{IPAC, California Institute of Technology, MC 100-22, 1200 E California Blvd, Pasadena, CA 91125, USA}%
\email{bomee@kasi.re.kr}%
\author[0000-0001-5382-6138]{Daniel~C.~Masters}%
\affiliation{IPAC, California Institute of Technology, MC 100-22, 1200 E California Blvd, Pasadena, CA 91125, USA}%
\email{dmasters@ipac.caltech.edu}%
\author[0000-0002-6025-0680]{Gary~J.~Melnick}%
\affiliation{Center for Astrophysics $|$ Harvard \& Smithsonian, Optical and Infrared Astronomy Division, Cambridge, MA 01238, USA}%
\email{gmelnick@cfa.harvard.edu}%
\author[0000-0002-8802-5581]{Jordan~Mirocha}%
\affiliation{Department of Physics, California Institute of Technology, 1200 E. California Boulevard, Pasadena, CA 91125, USA}%
\email{mirocha@caltech.edu}%
\author[0000-0001-9368-3186]{Chi~H.~Nguyen}%
\affiliation{Department of Physics, California Institute of Technology, 1200 E. California Boulevard, Pasadena, CA 91125, USA}%
\email{chnguyen@caltech.edu}%
\author[0000-0003-4408-0463]{Zafar~Rustamkulov}%
\affiliation{IPAC, California Institute of Technology, MC 100-22, 1200 E California Blvd, Pasadena, CA 91125, USA}%
\email{zafar@caltech.edu}%
\author[0000-0003-1841-2241]{Volker~Tolls}%
\affiliation{Center for Astrophysics $|$ Harvard \& Smithsonian, Optical and Infrared Astronomy Division, Cambridge, MA 01238, USA}%
\email{vtolls@cfa.harvard.edu}%
\author[0000-0003-4990-189X]{Michael~W.~Werner}%
\affiliation{Jet Propulsion Laboratory, California Institute of Technology, 4800 Oak Grove Drive, Pasadena, CA 91109, USA}%
\email{michael.w.werner@jpl.nasa.gov}%
\author[0000-0003-3078-2763]{Yujin~Yang}%
\affiliation{Korea Astronomy and Space Science Institute (KASI), 776 Daedeok-daero, Yuseong-gu, Daejeon 34055, Republic of Korea}%
\email{yyang@kasi.re.kr}%
\author[0000-0001-8253-1451]{Michael~Zemcov}%
\affiliation{School of Physics and Astronomy, Rochester Institute of Technology, 1 Lomb Memorial Dr., Rochester, NY 14623, USA}%
\affiliation{Jet Propulsion Laboratory, California Institute of Technology, 4800 Oak Grove Drive, Pasadena, CA 91109, USA}%
\email{mbzsps@rit.edu}

\begin{abstract}

We present map-making methodologies and preliminary spectral data cubes for \spherex, a NASA Explorer mission that launched in \launchMonth\ and has been performing an all-sky near-infrared spectral survey. The \spherex\ instrument observes from $\wlmin$ to $\wlmax~\um$ with a spectral resolution ranging from~$\Rmin$ to~$\Rmax$ and a pixel size of~$\pixsize$. We define a nominal set of \nchs~wavelength channels, each of which maps the entire sky approximately twice per year. Among the main mission goals is an investigation of the cosmic history of galaxy formation through intensity mapping of the extragalactic background light~(EBL), which is a primary motivation for the  map maker described in this work. The \spherex\ dataset contains a wealth of additional mapping targets, e.g., resolved galaxies and nebulae and diffuse clouds of Galactic dust and gas, which display strong spectral features such as hydrogen recombination lines, molecular-hydrogen lines and emission from polycyclic aromatic hydrocarbons~(PAHs). We describe how our map maker handles these various cases, how to mitigate foregrounds such as zodiacal light and upper-atmospheric emission and how to monitor and mitigate systematics and signal loss. Our maps are produced both in tangent-plane projection and in full-sky \healpix\ format. Specialized maps will be released to accompany future publications from the \spherex\ \st, and a public mosaic tool will be made available by the \irsaFull~(\irsa).

\end{abstract}


\section{Introduction}

Our aim in this work is to present preliminary map-making techniques for a variety of science goals that will be pursued by both the \spherex\footnote{Spectro-Photometer for the History of the universe, Epoch of Reionization and Ices Explorer.} \st\ and the broader community. The \spherex\ mission is an all-sky near-infrared spectral survey that is part of NASA's Medium-Class Explorers~(MIDEX) Program~\citep{Bock2026}. The satellite launched in \launchMonth\ and began science observations from low Earth orbit in May. The survey strategy covers the full sky approximately twice per year~\citep{Bryan2025}, and the survey first achieved complete coverage in late~2025. With some of the methods described in this work, the \spherex\ \st\ produced full-sky maps and released visualizations to the public; these include component maps that isolate emission from PAHs and regions of ionized hydrogen~(\HII).\footnote{\url{https://www.jpl.nasa.gov/images/pia26600-spherexs-first-all-sky-map/} \label{fn:allsky}}

Each scientific investigation is likely to prompt a specific set of improvements and refinements, but we expect the main processing steps to remain stable. The \spherex\ \st\ has constructed maps for studies of polycyclic aromatic hydrocarbons~\citep[PAHs,][]{Murgia2026}, ice absorption~\citep{Hora2026}, and the interstellar object \atlas~\citep{Lisse2026}. Several additional investigations are currently in progress, and a growing number are planned. The main map-making functionality will be available as part of a public mosaic tool on the website of the \irsaFull~\citep[\irsa,][]{Akeson2025}.\footnote{\url{https://irsa.ipac.caltech.edu/Missions/spherex.html} \label{fn:IRSA}}

The \spherex\ dataset enables the construction of maps in a nominal set of \nchs~wavelength channels spanning~$\wlmin$ to~$\wlmax~\um$. With a $\pixsize$~pixel size and a $\fov$~field of view, \spherex\ observations are complementary to those of the James Webb Space Telescope~(JWST), which operates in a similar wavelength range with an angular resolution that is much finer but a field of view that is much smaller~\citep{jwst2006}.

We provide a number of examples in this article, but we emphasize that these are intended only as illustrations of \spherex's map-making capabilities; further analysis is intended for future work. As examples of spectral data cubes, we present two complementary cases: one is the \catseye~(\secref{sec:mosaicens}), which is representative of small and compact targets, and the second is the full sky~(\secref{sec:fullsky_datacube}), which represents the opposite extreme.  In describing some techniques, we also present maps of the \heartsoul~(\secref{sec:coaddition}), the region surrounding the \bubble~(\secref{sec:selfconsistency}), the \tri~(\secs{\ref{sec:continuum_subtraction} and~\ref{sec:filter_considerations}}) and the \helix~(\secref{sec:specInterp}). As examples of scientific applications, we present spatially-resolved spectroscopy of the \catseye~(\secref{sec:spectroscopy}), a PAH~map of the \tri~(\secref{sec:continuum_subtraction}) and a spectrum of features correlated with interstellar dust~(\secref{sec:fullskycorr}). Validation standards will depend on the specific goals of each individual investigation; in this work, we describe several approaches, but none is to be considered universal.

\section{General Considerations}

We begin with a discussion of our general approach to \spherex\ map making. Some considerations are informed by the instrument and survey design; other choices are driven by software and algorithmic efficiency.

\subsection{Spectral Images \label{sec:spectral_images}}

A distinctive feature of \spherex\ observations is the use of linear variable filters~\citep[LVFs,][]{Korngut2026}. These filters are placed above the detectors and define the spectral passbands, which vary as a function of position within each image. Approximately, the transmitted wavelength changes along the $y$~axis of each detector array, and this is illustrated in \figref{fig:wavelengths} for all six of the \spherex\ detectors~(which are also referred to as ``bands''). 
\figenv{
    \includegraphics[width=\textwidth]{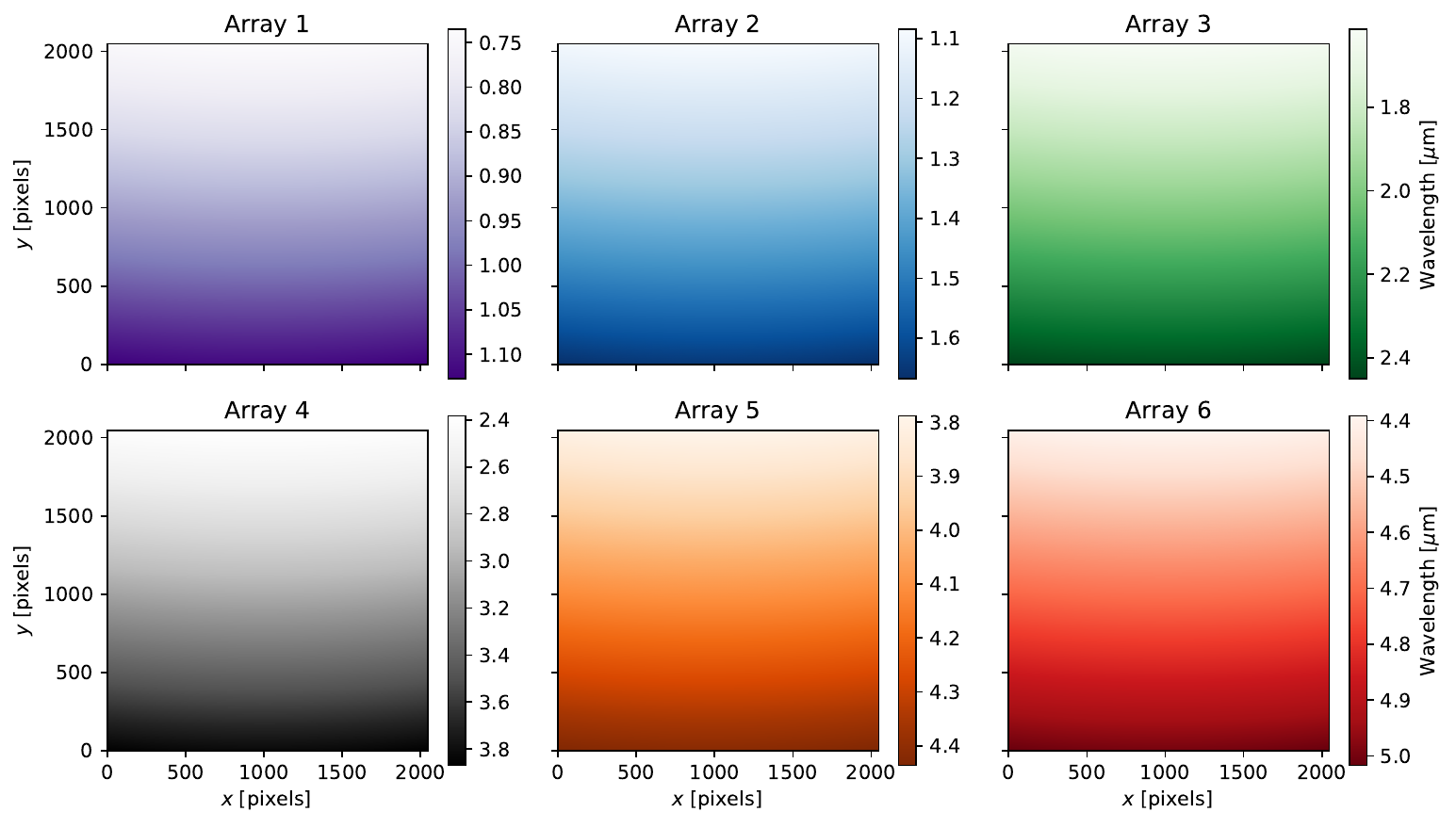}
    \caption{Measured central wavelengths of the six \spherex\ detector arrays~\citep{Hui2026}. Each detector array contains $\ndetside \times \ndetside$~optically active pixels. Wavelength increases in the negative-$y$ direction with a slight curvature due to the linear variable filters~(LVFs). The spectral resolution is~$\sim \Rsmall$ for arrays~1-4 and~$\sim \Rlarge$ for arrays~5 and~6.}
    \label{fig:wavelengths}
}
Although the central wavelength varies, the spectral resolution~$R$ is approximately constant within each~LVF. For detector arrays~1-4, we have $R \sim \Rsmall$; for detector arrays~5 and~6, $R \sim \Rlarge$. Each array consists of $\ndetside \times \ndetside$ optically active detector pixels, and there is substantial overlap in the passbands of neighboring pixels~\citep{Hui2026}. As projected onto the sky, the detector pixels are~$\pixsize$ on a side; since the optical~PSFs are comparable in width, the detector grid represents an undersampling, which was a design choice that enhances sensitivity to point sources~\citep{Korngut2026}. The effective image-level spatial resolution is, therefore, determined by a combination of the detector size and the optical PSFs. The six detector arrays are numbered~1 to~6, and wavelength increases nearly monotonically with array number; at the edges of consecutive arrays, there are small wavelength overlaps at the level of~$\order{0.01}~\um$.

The detector arrays are often divided into ``spectral channels'', which correspond to smaller wavelength ranges. Nominally, especially for purposes related to mission design and survey strategy, each array is divided into \nsubchs~spectral channels. Each nominal channel spans a wavelength range that is comparable to the LVF~passband widths~($\sim R \lambda$). In the representations of \figref{fig:wavelengths}, each channel corresponds roughly to a range of $y$~values. In \secref{sec:chdef}, we will consider non-standard channel definitions for the purposes of map making.

The LVFs impose vertical spectral gradients that cause an important entanglement between spatial and spectral coverage. During an exposure, all wavelengths are active, but different parts of the field of view are observed at different wavelengths. The survey strategy ensures that every line of sight is eventually observed by every spectral channel~\citep{Bryan2025}.

In \figref{fig:exp_panel}, we present an example of a single telescope pointing, which is centered near the \heart~(\heartcat).
\figenv{
    \includegraphics[width=\textwidth]{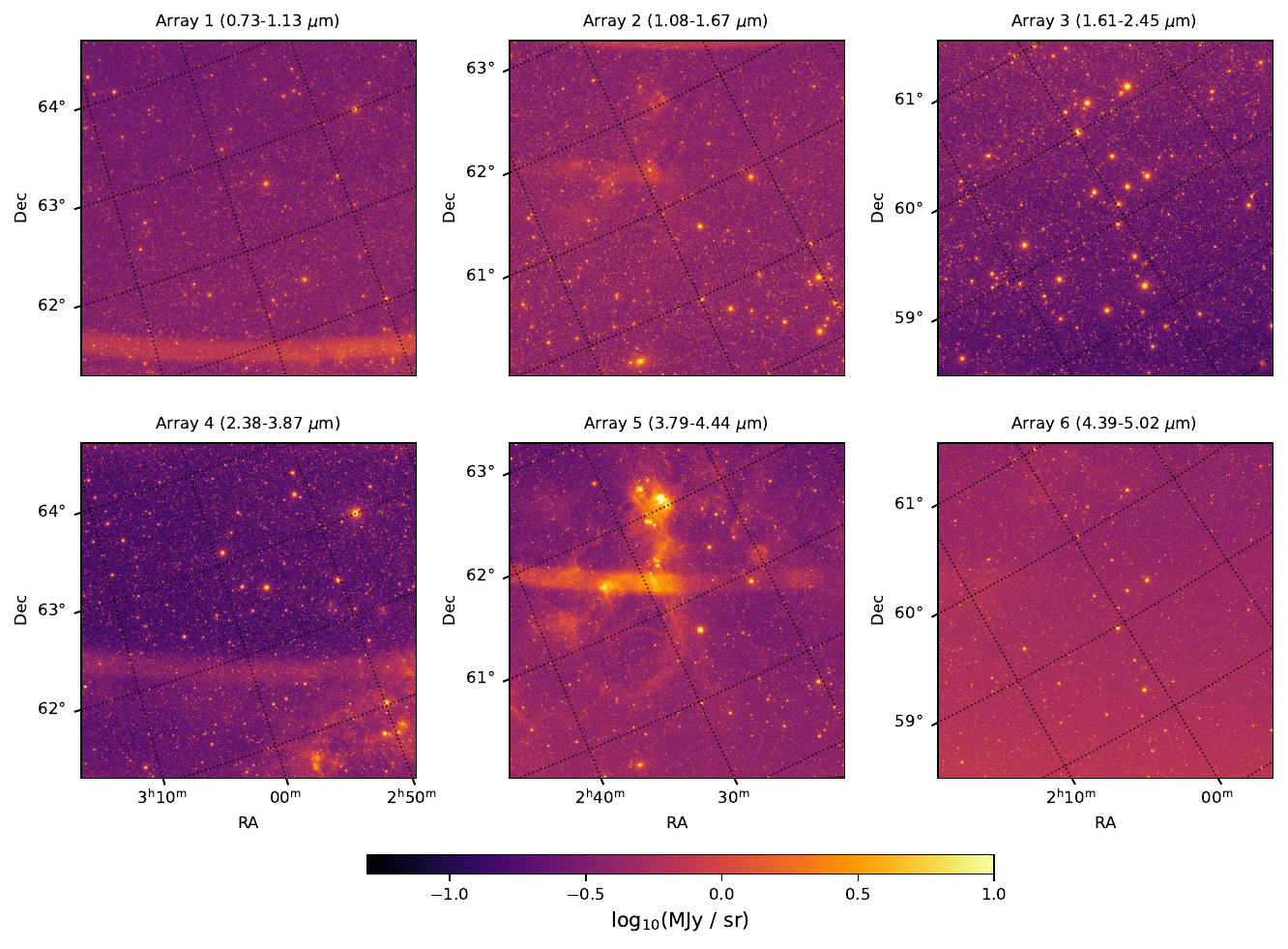}
    \caption{Spectral images from a single telescope pointing centered near the \heart~(\heartcat). Each panel's title indicates the detector array and, in parentheses, the associated wavelength range~(\figref{fig:wavelengths}). Wavelength increases in the downward direction across each array. The resulting images display a mixture of spectral and spatial features. Arrays~1, 2 and~3~(top row) are co-pointed, respectively, with arrays~4, 5 and~6~(bottom row).}
    \label{fig:exp_panel}
}
All six arrays operate simultaneously. For a given telescope pointing, images are collected from exposures lasting~$\exptime~\s$. A dichroic beam splitter~(DBS) sends shorter wavelengths to one focal plane and longer wavelengths to another. As a result, arrays~1, 2 and~3 are approximately co-pointed with arrays~4, 5 and~6, respectively; note the corresponding star patterns. The field of view of each image is~$\fov$. Combining arrays~1, 2 and~3~(or arrays~4, 5 and~6), the total instantaneous field of view is~$\fovfp \times \fov$. The \heart\ is visible in the center arrays~(2 and~5), but the spatial features are modulated by the wavelength dependence of the LVFs~(\figref{fig:wavelengths}).

For example, array~5 shows a bright horizontal stripe in \figref{fig:exp_panel}~(where the slight upward curvature, sometimes referred to as a ``smile'', is a well-characterized LVF feature that is due to the fabrication process). This particular stripe appears in the detector pixels that have significant transmission at~$\wlBra~\um$, which is the wavelength of the \Bra\ transition of neutral hydrogen~(\HI). Within the stripe, we are sensitive to the spatial features of \Bra\ emission; outside the stripe, we are mapping features that are different both spatially and spectrally.

Several similar but dimmer stripes can be identified in the other panels of \figref{fig:exp_panel}. Each is associated with the wavelength of a relatively prominent spectral feature. At the bottom of array~1, there is a stripe associated with the $\wlHe$-$\um$ transition of neutral helium~(\HeI). The \HeI~emission also appears as an enhancement near the upper edge of array~2, because there is a small wavelength overlap with array~1. In the upper half of array~2, there is a partially visible stripe associated with the \Pab\ transition of~\HI. In the lower half of array~4, the moderately bright stripe is associated with the $\wlPAH$-$\um$ emission feature of PAHs~\citep[e.g.,][]{Tielens2008}. Depending on the telescope pointing, other spectral features may become more or less prominent. In order to appear as a stripe, the emission structures must be large compared with the field of view. Smaller structures will induce brightness changes that are more localized.

Although \HeI\ emission is commonly observed in astrophysical structures, it was determined early in the observing campaign that the dominant signal at $\wlHe~\um$ is due to the upper atmosphere of the Earth~\citep{airglow2026}. The \spherex\ satellite is in low Earth orbit at an altitude of approximately~$\altitude~\km$. Although the vast majority of the Earth's atmosphere lies below \spherex, there is a non-negligible population of helium atoms at higher altitudes. This emission is referred to as ``helium airglow'' and depends on \spherex's orbital position as well as recent Solar activity. For the main science goals of \spherex, the helium airglow can be considered to be a contaminating foreground; it can also, however, serve as a spectral calibrator~\citep{Hui2026} and as an important observable in studies of the upper atmosphere~\citep{Kulkarni2025}. 

In addition to helium airglow, other atmospheric signals appear in \spherex\ data but at lower levels. These include airglow from oxygen, line emission from aurorae and shuttle glow produced by collisions between the spacecraft and upper-atmospheric particles~\citep{Bock2026,airglow2026}. These signals tend to be spatially smooth, which allows for the relatively simple filter that will be described in \secref{sec:contourFilter}. Spatial features are present and would need to be confronted for more effective foreground removal.

The smooth background glow in \figref{fig:exp_panel} is mainly due to zodiacal light, which is associated with interplanetary dust in the Solar System. For a preliminary \spherex-based measurement of the zodiacal-light spectrum, see \cite{Bock2026}. The minimum zodiacal-light signal occurs near~$\wlZodiMin~\um$, which is toward the bottom of array~4. In array~3, there is an overall vertical gradient that decreases toward longer wavelengths~(downwards); this is the falling spectrum of the scattering component of zodiacal light. In arrays~5 and~6, the gradients increase toward longer wavelengths; these are due to the Wien tail of the thermal component. In the other arrays, zodiacal light exhibits gradients that are too subtle to be seen clearly on the color scale of \figref{fig:exp_panel}.

\subsection{Survey Strategy \label{sec:scan}}

The \spherex\ telescope is relatively limited in the sky regions that it can observe at any given moment. The satellite is in a Sun-synchronous polar orbit that roughly follows the Earth's terminator~\citep{Bryan2025}. Over the course of a year, the plane of the orbit experiences a sidereal precession of~$360\degr$. Allowed pointings are constrained to avoid contamination from both sunlight and earthlight. Additional constraints have been imposed in order to mitigate shuttle glow and moonlight.

When accumulated over the course of a few days, \spherex's coverage tends to follow paths that are approximately constant in ecliptic longitude. On every orbit, \spherex\ observes near the ecliptic poles, where two ``deep fields'' have been defined on account of the relatively high number of repeated visits. Over the course of the year, the orbit precesses in ecliptic longitude and provides access to different parts of the sky. Deep-field observations continue throughout the year but with different detector orientations.

In \figref{fig:hits_panel}, we present full-sky coverage maps for six example channels with data collected as of \fullskydate.
\figenv{
    \includegraphics[width=\textwidth]{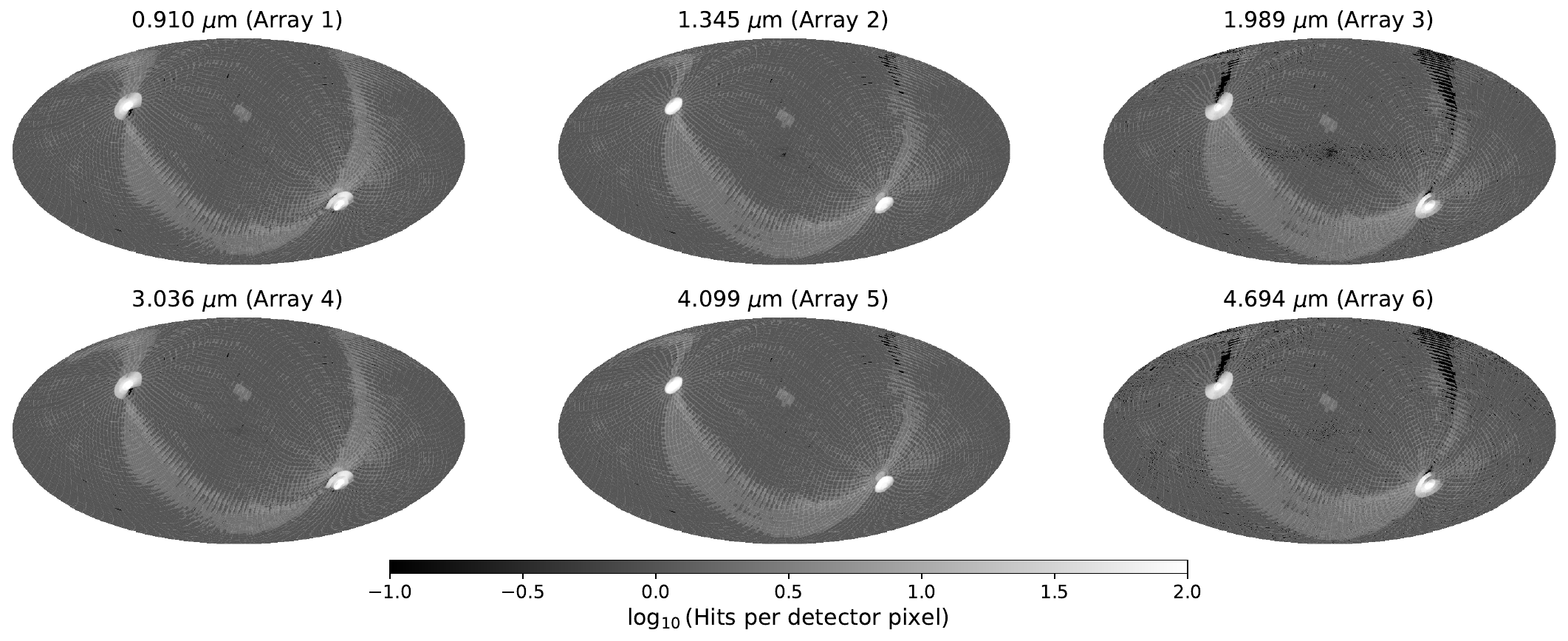}
    \caption{Coverage as of \fullskydate\ for one representative spectral channel from each of the six detector arrays. These maps are presented in Mollweide projection in Galactic coordinates with $\resolnsidehits$~pixels~($\nside=\nsidehits$). For each map pixel, we report the average number of hits in an area equal to a \spherex\ detector pixel~($\pixsize$). Due to the averaging, some map pixels can be assigned values below~1.}
    \label{fig:hits_panel}
}
Observation hits have been binned into a \healpix\ representation in Galactic coordinates with $\nside=\nsidehits$~\citep[][\secs{\ref{sec:reprojection} and~\ref{sec:fullskymaps}}]{Gorski2005}. These map pixels have an effective side length of~$\resolnsidehits$, which is much larger than \spherex's detector pixel size~($\pixsize$). The coarser map pixelization provides a convenient compression of the data and allows for the monitoring of large-scale coverage patterns. The \spherex\ survey strategy aims to observe the full sky with a resolution of approximately~$\pixsize$, so we renormalize each map pixel to report the average number of hits in an area equal to a \spherex\ detector pixel. This arithmetic can obscure some smaller-scale gaps in coverage, but those can be studied with finer resolution on smaller regions; several examples will be presented in \secs{\ref{sec:mapmaking}, \ref{sec:datacubes} and~\ref{sec:fullskymaps}}.

In \figref{fig:hits_panel}, the coverage maps are presented in Galactic coordinates, whereas the survey strategy is defined in ecliptic coordinates. The two deep fields can be easily identified, in the upper left and lower right, as the quasi-circular regions with significantly higher hits. The northern deep field~(upper left) is centered on the north ecliptic pole. The southern deep field~(lower right) is centered on ecliptic latitude~$\elatSDF$ and ecliptic longitude~$\elonSDF$, slightly offset from the south ecliptic pole in order to avoid the Large Magellanic Cloud~(LMC); the LMC is still observed but less frequently.

Outside of the deep fields, large-scale features stretch from one ecliptic pole to the other. These correspond approximately to paths of constant ecliptic longitude. Narrower features usually represent small overlaps between observations that were taken within a period of a few days. Toward the bottom left, we see elevated coverage over a much wider range of ecliptic longitudes; the orbit has precessed by more than~$180\degr$ and now allows for the reobservation of regions that were first visited approximately 6~months earlier. Because the survey follows ecliptic longitude only approximately, some regions have been reobserved while others remain unobserved. In the northern hemisphere, for example, arrays~3 and~6 display dark elongated streaks, which correspond to the last remaining gaps in coverage as of \fullskydate; as of the time of writing, however, these gaps have been closed.

Approximately $20\degr$~above the Galactic center, there is a small patch of elevated hits that can be found in all of the panels of \figref{fig:hits_panel}. The extra coverage is the result of a targeted campaign to observe interstellar object \atlas\ in August~2025~\citep{Lisse2025,Lisse2026}. Due to the time sensitivity of these measurements, the campaign was performed in a special observing mode.

The Galactic center was observed completely, but some of the images are currently missing adequate astrometric solutions. In \figref{fig:hits_panel}, array~3 shows the most noticeable deficit in coverage near the Galactic center, but the same phenomenon appears, to a lesser degree, for all arrays. As of the time of writing, the Level-2~(L2) fine-astrometry algorithms are more likely to fail in regions that are especially crowded with bright stars; affected images have been excluded from our maps. The fine-astrometry algorithms will be updated as part of the year-1 data release~\citep{Akeson2025}; these data are scheduled to become available in November~2026.

Although the general coverage patterns are similar for the spectral channels presented in \figref{fig:hits_panel}, there are clear differences. Note the variation in the sizes and shapes of the coverage of the deep fields. As noted above, arrays~3 and~6, which are co-pointed, are missing coverage in the northern Galactic hemisphere, whereas the other arrays have covered the sky almost completely but with slightly different patterns. In \figref{fig:hits_panel}, we have included only a single spectral channel per array; between neighboring channels, the differences are smaller.

\subsection{\skysim \label{sec:skysim}}

We make extensive use of the \spherex\ \skysim~\citep{Crill2025}, a software tool developed by the \spherex\ \st\ with the purpose of modeling data. The \skysim\ is able to produce mock images that contain the main instrumental and astrophysical contributions to \spherex\ data. Before launch, the \skysim\ was crucial for survey planning and analysis-pipeline development. After launch, the \skysim\ continues to guide scientific analyses by allowing for known signals, such as those presented in \secref{sec:simmaps}, to be propagated through pipeline algorithms.

The role and reliability of the \skysim\ depend on the particular needs of each individual scientific investigation. The \skysim\ continues to be updated to better imitate measured sky signals and in-flight instrument performance. For the map maker described in this article, the \skysim\ also enables the construction of source masks~(\secs{\ref{sec:pixel_masking} and~\ref{sec:sourceremoval}}) and deprojection templates~(\secs{\ref{sec:deproj} and~\ref{sec:fullskydeproj}}). The \skysim\ allows for data components to be studied independently or in superposition. 

In most \spherex\ images, the main data components are zodiacal light, starlight, diffuse Galactic light~(DGL), extragalactic background light~(EBL), photon noise, read noise and dark current. See \cite{Crill2025} for examples of simulated spectral images and for details of the underlying models. 

Zodiacal light, from a map-making perspective, represents a diffuse, time-variable foreground. Over the course of a year, the \spherex\ satellite travels through the zodiacal cloud, which has large-scale asymmetric structure~\citep[e.g.,][]{Kelsall1998,Wright1998}. In \figref{fig:exp_panel}, the zodiacal light appears as a smooth background gradient in each image; most of the gradient is due to the wavelength dependence of the LVFs~(\secref{sec:spectral_images}), but there is also lower-level spatial structure. Zodiacal-light simulations are helpful for monitoring foreground contamination~\citep[e.g.,][]{Murgia2026}, for testing image-level filtering schemes~(\secref{sec:contourFilter}) or for constructing map-level deprojection templates~(\secs{\ref{sec:deproj} and~\ref{sec:fullskydeproj}}).

Starlight is the brightest signal in \spherex\ data and acts as both a target and a foreground. Catalog-based starlight simulations are used in the construction of source masks for image-level filtering~(\secref{sec:contourFilter}) and smoothing~(\secref{sec:smoothing}) and for map-level deprojection~(\secs{\ref{sec:deproj} and~\ref{sec:fullskydeproj}}) and statistical analysis~(e.g., power-spectrum estimation). We simulate an image or a map of starlight and construct a mask by setting an intensity threshold for allowed leakage. The efficiency of the mask depends on the accuracy of the point-spread-function~(PSF) model and the cataloged source fluxes, both of which will improve over the course of the \spherex\ mission. Starlight simulations are also helpful as small-scale input signals that test pipeline algorithms~(e.g., \secref{sec:reprojection})

Diffuse Galactic light~(DGL), which is produced by starlight scattered from interstellar dust grains, is an important example of a celestial signal that is most significant on scales as large or larger than the \spherex\ field of view~(\secref{sec:filter_considerations}). By simulating DGL, we can test the reconstruction of large-scale features and characterize the impacts of map-making operations, e.g., the loss of information due to filtering or deprojection. The \skysim\ includes $3.3$-$\um$ PAH emission as part of what is considered DGL. As a prominent spectral feature, the simulated PAH signal allows for testing of continuum subtraction~(\secref{sec:continuum_subtraction}).

The extragalactic background light~(EBL), the integrated emission from all sources external to the \milkyway, is the main target of the \spherex\ investigation of the cosmic history of galaxy formation. Much of the EBL is integrated galactic light~(IGL), the accumulated emission from relatively bright galaxies that are individually cataloged and that play a role similar to that of stars in our development of map-making methodologies~(see above). Some nearby galaxies are resolved by \spherex, but most are point-like. We can generate mock EBL skies from a reference catalog of known objects measured by, e.g., \twomass~\citep{2MASS2006} and \wise~\citep{Wright2010}; in this case, the spectral energy distributions~(SEDs) are interpolated from the available photometry. We can also generate mock skies that are purely synthetic~\citep{Mirocha2026} in order to model galaxies over broad ranges of mass and redshift and to include their satellites and intrahalo light~(IHL). The EBL simulations are used to test sensitivity to different types of signals and to monitor distortions induced by the map-making process.

Photon noise can be simulated for all optical components. Over much of the sky, photon noise is dominated by the zodiacal light; the exceptions are at the locations of bright point sources and near the Galactic plane. In general, \spherex's raw sensitivity is primarily limited by photon noise, and the \skysim\ allows for photon-noise estimates that can be propagated through analysis operations.

Read noise is associated with the operation of the detectors and amplifiers and is subdominant to photon noise but not negligible. Because read noise is independent of incident optical signals, it produces different patterns in \spherex\ images and is subject to different statistical distributions. 

Dark current is the consistent signal that is measured in the detectors in the absence of optical stimulation. This produces a fixed pattern in all of the images recorded by each detector array. Because \spherex\ has no in-flight shutter, dark current cannot be measured directly but must be estimated as part of the gain-calibration algorithm~\citep{Akeson2025}. We can simulate dark current in order to test calibration algorithms and to assess the impact of mitigation strategies such as the subtraction of in-flight estimates.

Some simulation components must be coordinated even if they are studied independently. For example, photon noise depends on the assumed optical signals, and the fidelity of the photon-noise estimates is, therefore, dependent on the fidelity of the sky-signal modeling. If dust reddening is non-negligible, then the reddened components, e.g., starlight and~EBL, must be coordinated with~DGL. 

\subsection{Linearity \label{sec:linearity}}

As a map-making design principle, we aim for each operation to be linear in the sense that, at any point in the process, we can represent the data as a superposition of components. For example, consider the case of $n$~simulation components, where the components may represent stars, DGL, read noise, etc. For both images and maps, we could represent the data vectors as $\vec{x} = \sum_{i=1}^n \vec{x}_i$, where $\vec{x}_i$~is the $i$th~constituent component. If we act on the data with a linear transformation~$T$, we have
\begin{equation}
\label{eq:linearity}
\vec{y} \equiv T(\vec{x}) = \sum_{i=1}^n T(\vec{x}_i) = \sum_{i=1}^n \vec{y}_i ,
\end{equation}
where $\vec{y}$ and $\vec{y}_i$~represent the transformations of~$\vec{x}$ and~$\vec{x}_i$, respectively. For each exposure, we can form a single $n$-component image~($\vec{x}$), or we can form $n$~single-component images~($\vec{x}_i$). We can propagate these images through our map maker~($T$) and produce either a single $n$-component map~($\vec{y}$) or $n$~single-component maps~($\vec{y}_i$). If all operations are linear, the $n$-component map will be equal to the sum of the $n$~single-component maps~(\eqref{eq:linearity}). In \secref{sec:simmaps}, we will present an example of superposition with $n=7$.

An example of a linear operation is our simulation-based source masking~(\secref{sec:skysim}). Each source mask is constructed independently of the data to which it is applied. In simulation, we can apply the same source mask to each data component and study each individually. We can study how the mask affects the sources that it is meant to target, but we can also study how the mask distorts or obscures components such as zodiacal light or~DGL.

Another important example is that of image-level filtering~(\secref{sec:contourFilter}), which is mainly targeting zodiacal light. The real images contain zodiacal light in combination with the other main components described in \secref{sec:skysim}. If our map maker is linear, we can study the zodiacal-light component by itself in order to evaluate the efficacy of our image-level filter. At the same time, we might wish to monitor the impact of the filter on target observables such as the~EBL; in general, the filter will suppress or distort features in the target signal. Additionally, we can study the contaminants from, e.g., unmasked starlight that will also be affected by the filter.

Each image is reprojected~(\secref{sec:reprojection}) to a map grid, and other images are then coadded~(\secref{sec:coaddition}) to build the full map and increase sensitivity. These processes are linearly associative, which allows for the map-making process to be subdivided by time or by wavelength range. At any point, the subdivisions can be summed to form the map for a longer time period or for a larger wavelength range. 

\section{Image Processing \label{sec:image_proc}}

We now describe the operations that are applied to each image before it is reprojected and coadded to the map. Additionally, some images are rejected completely, mostly on account of excessive transient signals or a failure to find an astrometric solution.

Figure~\ref{fig:filter_panel} illustrates the main image-processing steps that will be described in this section. 
\figenv{
    \includegraphics[width=\textwidth]{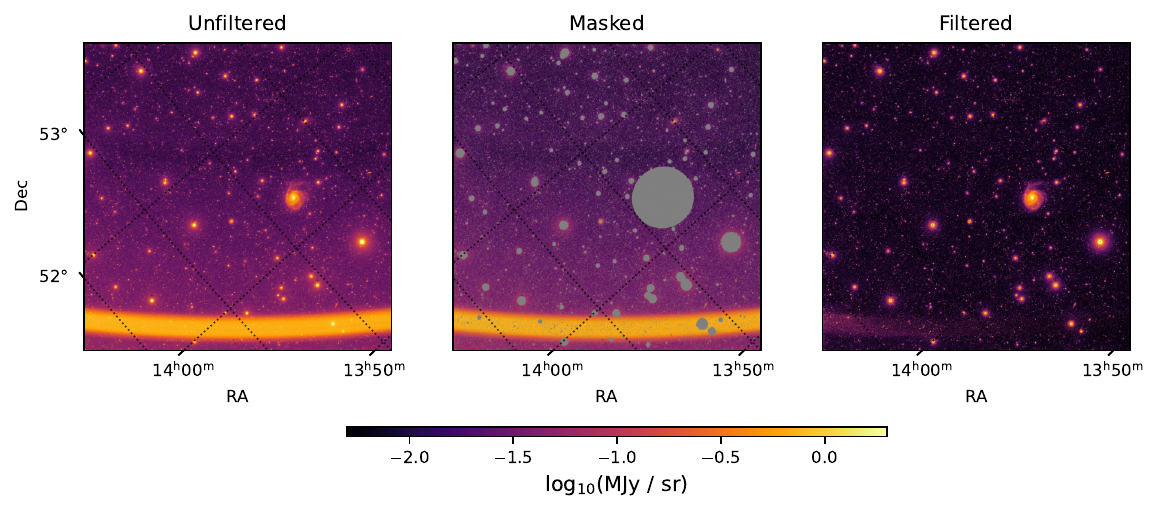}
    \caption{Different stages of image processing for an example exposure from array~1. The \pinwheel~(\pinwheelcat) can be seen right of center in each panel, and the bright band at the bottom is helium airglow~(\secref{sec:spectral_images}). (\emph{Left})~An unprocessed L2~image~(\secref{sec:L2}). We have subtracted an offset~($0.135~\MJysr$) in order to emphasize the spectral and spatial structure of the zodiacal light. (\emph{Middle})~The same image with a mask to suppress flagged pixels, point sources and~\pinwheelcat~(\secref{sec:pixel_masking}). (\emph{Right})~The same exposure but now filtered to remove zodiacal light and airglow~(\secref{sec:contourFilter}). The source and object masks have been lifted, but the flag mask is retained. In the filtered image, some of the residuals are negative, which cannot be seen on this logarithmic color scale. The negative residuals must be checked separately to ensure that they are small compared with the target signals. Notice that the helium airglow has been significantly reduced but not completely.}
    \label{fig:filter_panel}
}
For this example, we have chosen a pointing near the \pinwheel~(\pinwheelcat). Unlike the \heart~(\figref{fig:exp_panel}), \pinwheelcat\ is small compared with the \spherex\ field of view, so it can be efficiently masked during the filtering step. We have lowered the minimum of the color scale in \figref{fig:filter_panel} in order to emphasize the impact of zodiacal light and atmospheric emission and in order to extract finer features from the processed image.

\subsection{Level-2 Images \label{sec:L2}}

In general, our map maker operates on the Level-2~(L2) images that are described in \cite{Akeson2025} and that are produced by the \ssdcFull~(\ssdc). These images are publicly available on the \irsa~website~(\fnref{fn:IRSA}). Figure~\ref{fig:exp_panel} presents a panel of L2~images for a single telescope pointing near the \heart. For this section, however, the guiding example is in \figref{fig:filter_panel}, which displays a pointing near~\pinwheelcat; the leftmost panel is an unprocessed L2~image.

The L2~images include an astrometric solution with a figure of merit to indicate alignment quality. For the map maker described in this article, we typically restrict to the images that achieve the highest astrometry rating~(alignment~$< 1\arcsec$). When astrometric precision is less important, e.g., when making maps of diffuse structures on coarse grids, we may permit the usage of images with lower ratings. 

Flags are provided to indicate pixels that are unreliable. Some of these pixels are known to be nonfunctional. Some are flagged due to on-board detection of transients or other anomalous signals. Very bright signals, e.g., at the centers of stars, are often flagged due to detector nonlinearity. These flags contribute to the construction of image-level pixel masks~(\secref{sec:pixel_masking}).

Calibrations have been applied based on templates derived from Level-1~(L1) images. An estimate of dark current has been subtracted, and the images have been flat-fielded, i.e., corrected for relative gain fluctuations. The L2~pipeline also applies an absolute gain calibration, which converts the images from photocurrent~(in units of~$\mathrm{e}^{-}\mathrm{s}^{-1}$) to specific intensity~($\MJysr$).

\subsection{Pixel Masking \label{sec:pixel_masking}}

At the image level, we apply two types of masks. First, we often wish to remove pixels that are unreliable or inappropriate for the intended science goal; these pixels will be ignored for all subsequent map-making steps. Second, if we are performing image-level filtering, there are pixels that we wish to mask in order to better isolate the target foregrounds, e.g., zodiacal light and atmospheric emission; this mask is temporary and will be lifted before reprojection and coaddition.

In \figref{fig:filter_panel}, the middle panel shows the original L2~image with a mask informed by pixel flags~(\secref{sec:L2}), point-source simulations~(\secref{sec:skysim}) and the geometry of~\pinwheelcat. In cases for which outlier pixels persist, we can optionally apply sigma clipping to the source- and flag-masked image; within narrow wavelength ranges, we remove pixels that deviate significantly from the median. To maintain linearity~(\secref{sec:linearity}), we base each sigma-clipping mask on the real data; when handling a simulated image, we use the same mask as for the corresponding real image.

The L2~images~(\secref{sec:L2}) provide pixel flags for various types of detector behavior, and our image-level pixel mask can optionally select any subset of these flags. Typically, we mask pixels that have been flagged for transients, non-functionality, irrecoverable saturation, strong image persistence and other similar pathologies. To maintain linearity~(\secref{sec:linearity}), we build the flag mask from the real data; for simulated images, we apply the flag mask from the corresponding real images. Our pixel-flag mask is retained after filtering, so it removes information from the map-making process; it may sometimes be desirable, especially in the case of bright signals, to be more tolerant to certain types of flagged pixels. After filtering, when the source mask has been lifted, some of the flags are noticeable as missing data points in the centers of bright stars.

For filtering, we must mask stars and galaxies that are bright relative to zodiacal light and atmospheric emission. We form source masks by simulating source images~(\secref{sec:skysim}) and setting intensity thresholds. The intensity threshold balances source masking with sky area. In crowded regions, a strict threshold could cause the entire image to be masked. At relatively large Galactic latitudes, such as the location of~\pinwheelcat~($b \approx 60\degr$, \figref{fig:filter_panel}), we typically set the intensity threshold to~$\order{0.01}~\MJysr$, which allows zodiacal light to dominate while still retaining enough sky area for filtering operations. For the particular example of \figref{fig:filter_panel}, we set the threshold to~$0.01~\MJysr$ for dim sources~($\mAB > 10$) and~$0.002~\MJysr$ for bright sources~($\mAB \leq 10$); the brighter sources are masked more aggressively in order to suppress large-scale features from the extended PSFs, which are stronger in the real data than in the pre-launch optical simulations that inform the mask.

For extended objects, e.g.,~\pinwheelcat\ in \figref{fig:filter_panel}, we apply a circular mask of a specified radius. If the object is a scientific target, this mask will help to preserve the object's internal structure and avoid filter-induced distortions. This type of mask can be applied only to objects that are smaller than the \spherex\ field of view~(\secref{sec:filter_considerations}). After filtering, the mask is lifted.

\subsection{Contour Filtering \label{sec:contourFilter}}

We optionally apply an image-level filter to remove zodiacal light and atmospheric emission. This filter targets any component that is spatially smooth. Since \spherex\ observations are made with~LVFs~(\secref{sec:spectral_images}), a spatially smooth component will manifest as a vertical gradient that is determined by its~SED. The LVF~wavelength contours have a slight upward curvature~(``smile''), which is visible in \figs{\ref{fig:wavelengths}, \ref{fig:exp_panel} and~\ref{fig:filter_panel}}. We implement our filter by dividing each image into small wavelength bins and subtracting the mean value from each independently. The number of bins can be varied; smaller bins are more spectrally pure but contain fewer pixels. Typically, we form wavelength bins that span roughly 1~pixel in the vertical direction. Because these bins follow the LVFs' wavelength contours, we refer to this operation as a ``contour filter''.

The contour filter removes any component that is smooth within each wavelength bin on the scale of the \spherex\ field of view. Before filtering, we apply a pixel mask~(\secref{sec:pixel_masking}) to remove contributions that would bias the filter. After filtering, the mask can be lifted or altered.

The process is illustrated in \figref{fig:filter_panel}. Notice that the filter has heavily suppressed both the zodiacal light and the helium airglow. With the given color limits, the zodiacal-light residuals are nearly imperceptible, while the airglow residual shows horizontal spatial structure. The contour filter removes only a mean value, so any smaller-scale features will persist. Nevertheless, the airglow has been significantly suppressed, and we demonstrate the scientific capabilities of the airglow wavelength~($\wlHe~\um$) in \figs{\ref{fig:datacube}, \ref{fig:aperturephotometry} and~\ref{fig:filtcoadd_panel}}.

Because the contour filter is essentially a mean subtraction, it acts linearly on the constituent components of each image~(\secref{sec:linearity}). When studying simulated images, we can filter each component separately or together. Although the filter suppresses unwanted foregrounds, it also removes information from target signals, and this effect must be monitored.

With a sufficiently reliable model of zodiacal light, a template-based filter could be implemented. The \skysim\ provides a model that may be adequate, especially if an additional wavelength-dependent scaling is allowed, but it does not, as of the time of writing, account for atmospheric signals such as airglow, shuttle glow and aurorae. For this reason, we tend to prefer the contour filter, which is more aggressive than a template-based filter but also more robust.

\section{Map Making \label{sec:mapmaking}}

We now describe the process of combining \spherex\ images to produce maps. As for image processing~(\secref{sec:image_proc}), map making involves a number of choices that depend on the application of interest. There is no unique map-making prescription. In the coaddition process, the exposure contributions are approximately tiled, so we often refer to these maps as ``mosaics''.

\subsection{Channel Definitions \label{sec:chdef}}

We begin by dividing each detector array~(\secref{sec:spectral_images}) into wavelength ranges that we term ``spectral channels''. The survey plan~(\secref{sec:scan}) was designed under the assumption that each array would be divided into \nsubchs~channels. When the telescope is commanded to perform a short slew, it moves by approximately the width of one of these nominal channels. The survey plan ensures that each nominal channel observes every $\pixsize$~sky pixel approximately twice per year. If the arrays were divided into channels with smaller wavelength ranges, the maps would be more spectrally pure but have larger coverage gaps. In year~2, the survey strategy offsets the scans by half the width of a nominal channel, so the accumulated dataset could be divided into \nsubchsDouble~channels per array without loss of coverage. For a given sky area, we make a separate map for each spectral channel.

In \figref{fig:ChDef}, we present three examples of channel definitions for array~1.
\figenv{
    \includegraphics[width=\textwidth]{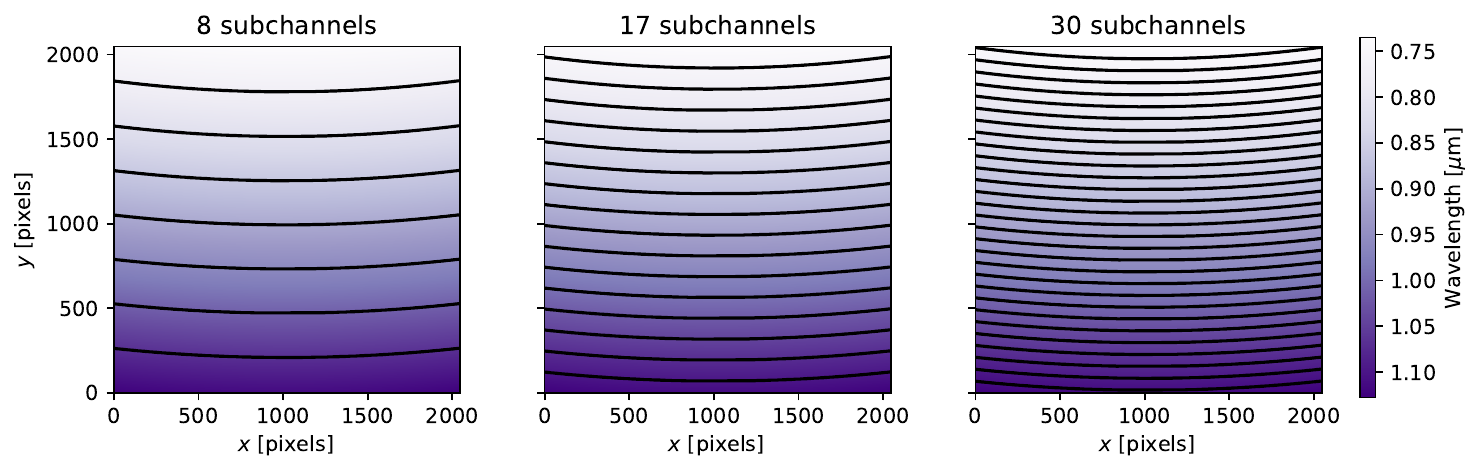}
    \caption{Three examples of channel divisions for array~1. The middle panel is similar to the definition used for survey planning~(\secref{sec:scan}). In the left panel, the wavelength ranges are wider, which increases spatial coverage per channel. In the right panel, the channels are more spectrally pure, but spatial coverage may contain gaps depending on the time period of observations that have been included.}
    \label{fig:ChDef}
}
For other arrays, the divisions are similar. For the purpose of survey planning~(\secref{sec:scan}), the channels are defined geometrically, since the goal is to determine the telescope movements. For the map maker described in this article, the channels are defined relative to the measured spectral responses of the \spherex\ detectors~\citep{Hui2026}. We find the minimum and maximum wavelengths and divide the entire array into logarithmically spaced wavelength bins. Because of the curvature of the LVF response, the edge channels have shapes that are different from those in the interior of the array; this can be seen in all of the examples of \figref{fig:ChDef}. Unless otherwise stated, we assume that the number of channels per array is~\nsubchs.

\subsection{Reprojection \label{sec:reprojection}}

We reproject from each image to a mosaic grid. The mosaic grid is defined in either celestial or Galactic coordinates as a tangent-plane projection~(though we will also, in \secref{sec:fullskymaps}, consider a \healpix\ representation for full-sky mapping). The mosaic pixel size is typically chosen to be similar to the size of \spherex's detector pixels~($\pixsize$). In regions of high coverage, e.g., in the deep fields~(\secref{sec:scan}), the mosaic pixels are often chosen to be finer. In all cases, reprojection causes an additional distortion of sky signals, especially at small scales, and this can be characterized with simulations~(\secref{sec:skysim}) when relevant for a given science goal.

We consider two reprojection algorithms. Our preferred algorithm is a simple binning of the image pixels into the mosaic grid. This is implemented as an intensity-weighted 2D~histogram of the image-pixel positions. In this binning scheme, each image pixel contributes to only a single mosaic pixel. Another method is a linear interpolation from the image pixels to the mosaic grid; interpolation is helpful when finer resolution is desired despite limited coverage. The binning procedure is faster and preserves more information, but it is more constrained in the choice of mosaic grid, especially when coverage is light. In \secref{sec:deep_fields}, the tradeoffs are demonstrated for the case of the deep fields. Unless otherwise stated, we assume that reprojection has been performed by pixel binning.

Reprojection can also be performed by geometrically partitioning each image pixel into the overlapping mosaic pixels. Variable-Pixel Linear Reconstruction~\citep{Fructher2002}, also known as ``drizzling'', is a generalization that allows for the image pixels to be shrunk into smaller ``drops'' before the overlapping areas are computed. We have found that area-based reprojection is relatively slow and that image-level information is best preserved in the case that the drop size is infinitesimally small, which is equivalent to the binning method described above.

\subsection{Coaddition \label{sec:coaddition}}

Each image is coadded to the mosaic grid, and we gradually build up spatial coverage. At the same time, we track the number of contributions~(hits) to each mosaic pixel, and we track the central wavelengths of those contributions~(\secref{sec:spectral_images}).

In \figref{fig:coadd_panel}, we present three stages in the process of coaddition for a region containing the \heartsoul~(\heartcat\ and \soulcat), which also featured in the exposure panel of \figref{fig:exp_panel}.
\figenv{
    \includegraphics[width=\textwidth]{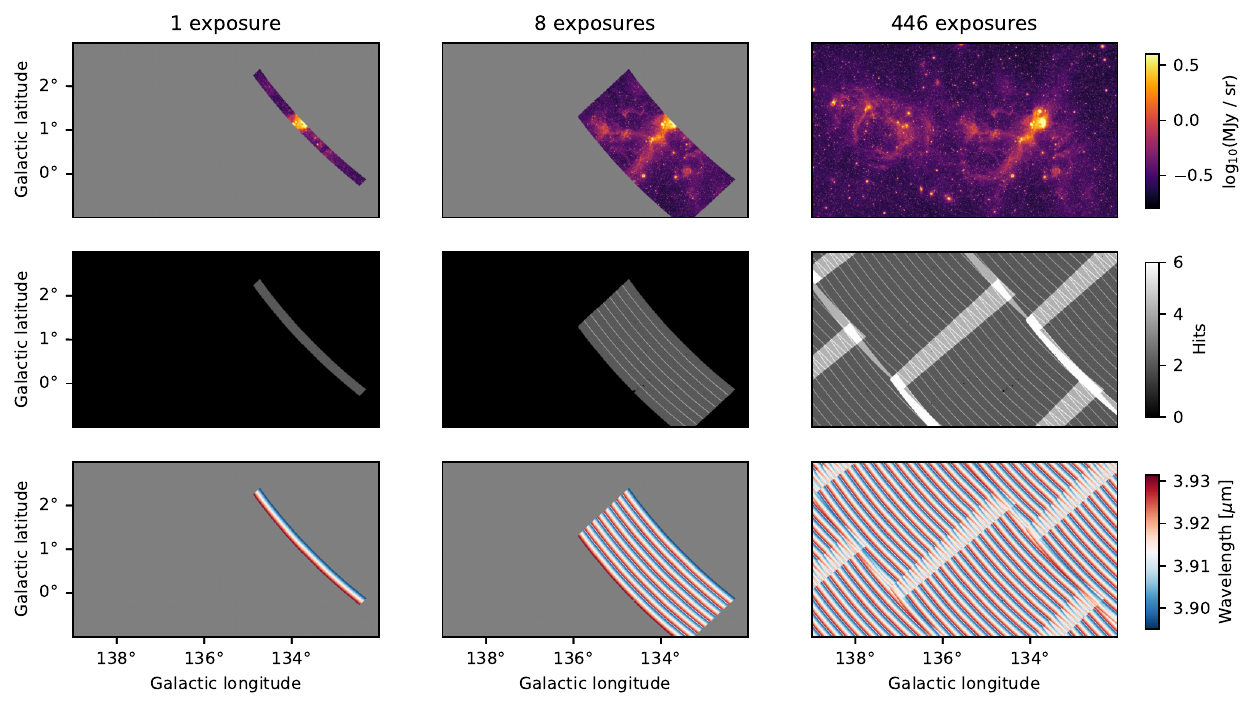}
    \caption{Three stages of coaddition for a region covering the \heartsoul~(right and left sides, respectively, of each panel). The mosaic grid is $\dimsCoadd$ with $\pixsizeCoadd$~pixels. Each column has a different number of contributing exposures. These maps are made for a spectral channel centered on~$\wlCoadd~\um$, which is in array~5. (\emph{Top row})~Maps produced by the three example stages of coaddition. (\emph{Middle row})~Hits maps showing how many detectors contributed to each mosaic pixel. (\emph{Bottom row})~Detector wavelengths that contributed to each mosaic pixel. In regions of repeated coverage, the wavelengths tend to average to a central value~(white).}
    \label{fig:coadd_panel}
}
For this example, the image processing~(\secref{sec:image_proc}) consists only of a pixel-flag mask. Zodiacal light and atmospheric emission are still present in the data, but they are subdominant. An image-level filter would remove information on scales similar to the \spherex\ field of view~($\fov$), which is smaller than the example mosaic grid~($\dimsCoadd$); so we skip filtering for this case. We consider a single spectral channel centered on~$\wlCoadd~\um$. We have deliberately avoided strong spectral features, e.g., hydrogen lines or the $\wlPAH$-$\um$~PAH~emission feature, which create additional complications that will be discussed in \secref{sec:specInterp}. The pixel size is chosen to be~$\pixsizeCoadd$, which avoids coverage gaps with the binning reprojection described in \secref{sec:reprojection}.

In \figref{fig:coadd_panel}, each column presents the mosaic with a different number of contributing exposures. In the leftmost column, we see how an individual exposure is reprojected. Our example channel is in the upper~(lower-wavelength) portion of array~5 but not on the edge, so we see the characteristic curvature of the LVF~response. The middle row presents the number of hits in each mosaic pixel. When reprojection is performed by binning~(\secref{sec:reprojection}), as in this case, the hits depend on the relative sizes and orientations of the detector and mosaic grids. The bottom row presents the central wavelengths of the detectors that contributed to each mosaic pixel. We see the LVF~wavelength gradient in the reprojection of our example channel. 

As more exposures contribute, the spatial coverage expands~(middle and right columns of \figref{fig:coadd_panel}). The hits maps show regions of overlapping coverage. In the middle column, only a small number of exposures contribute, and they are all drawn from a relatively short time period. Notice that the pointing steps are coordinated with the width of the spectral channel, so there is only a slight overlap between neighboring exposures. With full coverage, the overlap regions are more significant, and the detector wavelengths tend to average to the central value. Notice the entanglement between spectral and spatial coverage.

\subsection{Self-consistency \label{sec:selfconsistency}}

As this manuscript was under development, \spherex\ began to reobserve some regions that lie outside of the deep fields. Some of the doubled coverage can be seen in \figref{fig:hits_panel}, which uses data through \fullskydate, but the reobserved area has expanded since then. With two complete passes, we can form a null test to check the self-consistency of our map-making procedures and to estimate map-level variance including but not limited to noise.

Our example null test is depicted in \figref{fig:diff_panel}, which is centered near the \bubble~(\bubblecat). 
\figenv{
    \includegraphics[width=\textwidth]{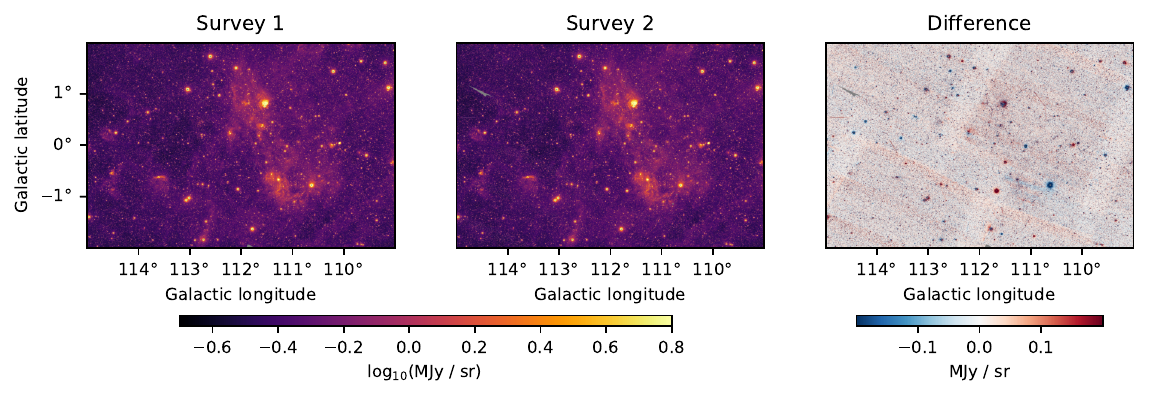}
    \caption{Null test from a comparison of surveys~1 and~2. The selected region is centered near the \bubble, which is the small~($\sim 15\arcmin$) bright object at $(l,b) = (112\fdg24,0\fdg24)$. These maps are made for a spectral channel centered on~$\wlDiff~\um$. (\emph{Left})~Map constructed entirely from survey~1. (\emph{Middle})~Map constructed entirely from survey~2. (\emph{Right})~Difference map. The color scale is symmetric with the maximum absolute value corresponding to the minimum of the scale that is used for the single-survey maps. Most of the differences would be imperceptible in the left two panels.}
    \label{fig:diff_panel}
}
At the time of writing, the \heartsoul~(\figref{fig:coadd_panel}) has been visited only once, but the region near the \bubble\ is similar; both contain diffuse structures that are comparable in size to the \spherex\ field of view. We constructed a map using images from survey~1 only, and then we separately made the same map using only survey~2. In the rightmost panel, we present the difference. The most significant residuals are at the locations of bright sources; if those sources are scientific targets, then it should be checked separately that the residuals are small fractionally. We focus instead on the larger-scale features, which have mostly canceled out in the difference. The median has decreased from~$0.431$~(survey~1) and~$0.435$~(survey~2) to~$0.003~\MJysr$, and the median absolute deviation from~$0.077$ and~$0.078$ to~$0.036~\MJysr$. In the difference map, $93.4\%$~of pixels are within the color limits~($\pm 0.2~\MJysr$). Some of the residuals show patterns that are characteristic of the scan strategy~(cf.~middle panel of \figref{fig:coadd_panel}); within a single survey, there are coverage overlaps that achieve reduced noise, and we see that in the difference. There are also features that resemble single-exposure reprojections of spectral channels; these may be due to time variability in the zodiacal light or atmospheric emission. We also find slender, linear features that are likely due to unflagged emission from Earth-orbiting satellites. As \spherex\ accumulates repeated coverage over the full sky, any region will be available for a null test like that of \figref{fig:diff_panel}, and the residuals can be compared to the required precision for the scientific application of interest.

For the null test in \figref{fig:diff_panel}, we have deliberately chosen a wavelength that avoids the sharpest spectral features in this region, though there may be some contribution from the 0-0~S(11)~transition of molecular hydrogen~(\Hmol,~$4.181~\um$). Recall from, e.g., the bottom row of \figref{fig:coadd_panel}, that there are small spatial variations in the observation wavelengths. Near a sharp spectral feature, e.g., an emission line, spatial and spectral features will be entangled and more sensitive to the exact distribution of observed wavelengths~(\secref{sec:specInterp}). In that case, a null test would need to account for the wavelength differences between the two surveys.

\subsection{Deep Fields \label{sec:deep_fields}}

In the deep fields~(\secref{sec:scan}), many repeated visits are made with different detector orientations and offsets. This allows for greater sensitivity and also for finer mosaic pixels. Small structures will still be suppressed at scales smaller than \spherex's detector pixels~($\pixsize$), but they can be partially recovered with the additional information that is available in the deep fields. Note that, with a smaller mosaic pixel, fewer image pixels contribute to each, and this will cause an increase in mosaic-level pixel noise.

In the deep fields, we typically set the mosaic pixel size to~$\pixsizeDeep$, and we find that finer structures are better recovered with binning reprojection than with interpolation~(\secref{sec:reprojection}). Consider the example mosaics in \figref{fig:reproj_demo}, which shows a small region~($\dimsReproj$) of simulated stars~(\secref{sec:skysim}) in the north deep field with $\pixsizeDeep$~mosaic pixels.
\figenv{
    \plotone{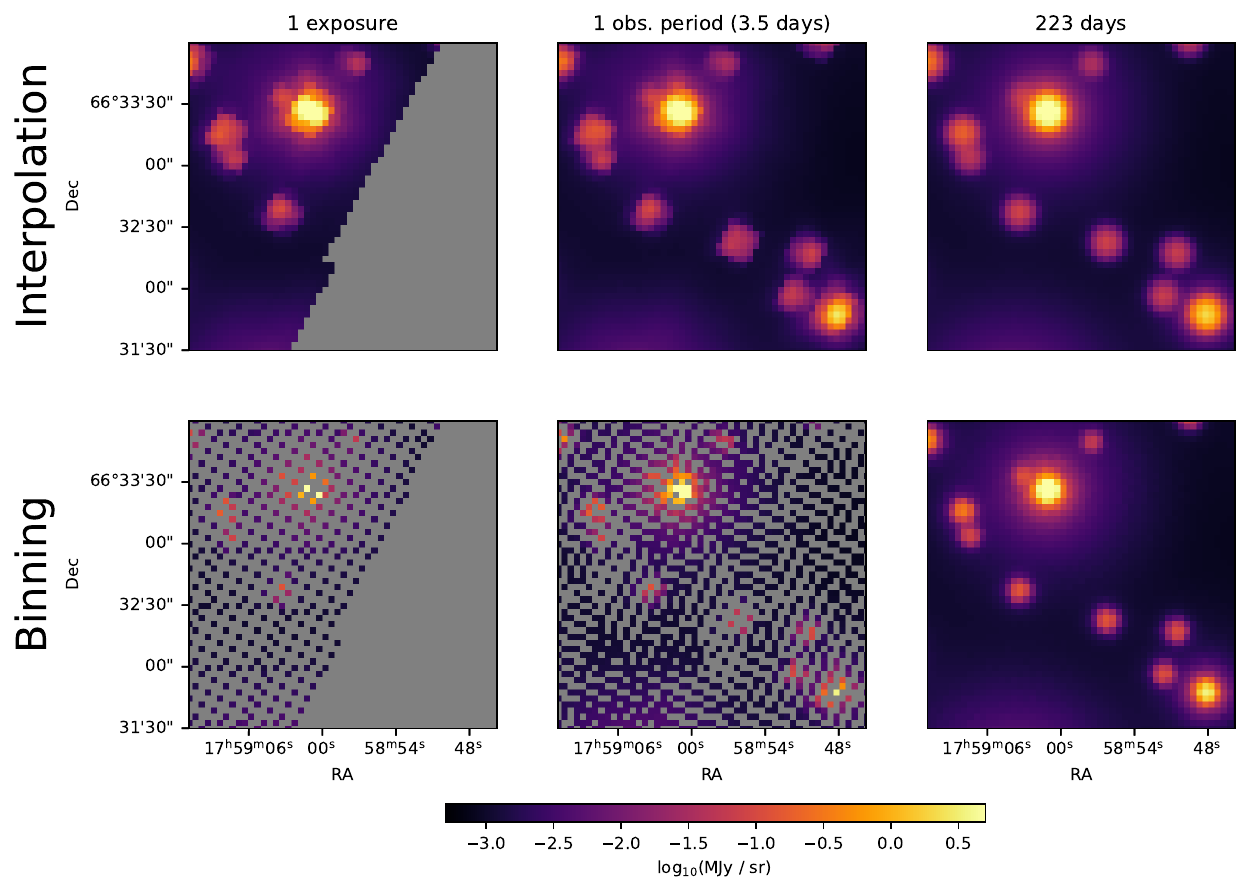}
    \caption{Comparison of reprojection algorithms~(\secref{sec:reprojection}) in an area of~$\dimsReproj$ in the north deep field~(\secref{sec:scan}) with a mosaic pixel size of~$3''$. To demonstrate the differences for small-scale features, we use simulated stars~(\secref{sec:skysim}). Each column presents a mosaic with a different number of exposure contributions. Where interpolation~(top row) allows for contiguous coverage with a smaller number of exposures, the binning method~(bottom row) ultimately produces stars that are sharper.}
    \label{fig:reproj_demo}
}
We present three stages of coaddition with the two different reprojection algorithms described in \secref{sec:reprojection}. With a single exposure, interpolation is superior, because it allows for contiguous coverage on the mosaic grid; notice, however, that the stars are rendered with features that are subtly rectilinear. The binning reprojection yields many gaps, because the detector pixels are significantly larger than the mosaic pixels; the information content, however, is less distorted, because the detector contributions are closer to their original positions. Both methods are subject to PSF~convolution, which is a feature of the telescope optics rather than the data processing. As more exposures are coadded, the interpolated mosaic symmetrizes the stars, and the binned mosaic fills in the gaps. After \daysReproj~days~(right column in \figref{fig:reproj_demo}), both mosaics have complete coverage, but the binning method yields stars that are sharper. Whereas the interpolation spreads starlight with every exposure, the binning method preserves more of the image-level information and produces a result that is closer to a convolution of the star field with the azimuthally-symmetrized detector pixel. 

We can compare the reprojected results to a simulation of the stars convolved with the optical PSF, i.e., a simulation that bypasses all of the map-making steps; both reprojection methods widen the sources relative to the optical PSF, but the binning method widens less. Because \spherex\ PSFs are undersampled, especially at the shortest wavelengths, a preferable resolution metric is the number of effective noise pixels~$\neff \equiv 1/\sum_{i}{p_i^2}$, where $p_i$~is the fraction of a single source's flux in pixel~$i$~\citep{Korngut2026,Cutri2012}. With simulations like those presented in \figref{fig:reproj_demo}, we can compute~$\neff$ for interpolation and binning, and we can also compute~$\neff$ for a simulation that bypasses map making and includes only the effects of the optical~PSF. As $\neff$~is a measure of solid angle, we can use~$\sqrt{\neff}$ as a measure of PSF width. For the examples shown in \figref{fig:reproj_demo}, we used a channel centered on~$0.910~\um$. Relative to the optical~PSF, we find that binning increases the effective PSF width by a factor of~$\sim 1.6$; interpolation increases the width by~$\sim 2.2$. Much of the widening is due to the $\pixsize$~detector pixels~(\secref{sec:spectral_images}), which are substantially larger than the optical~PSF for \spherex's shortest wavelengths. This effect is weaker for wavelengths that are longer. For example, if we repeat these calculations for a channel centered on~$4.694~\um$, we find that binning increases the effective PSF width by~$\sim 1.3$; interpolation by~$\sim 1.6$. In all cases, interpolation widens the PSF by a larger factor. For these reasons, we prefer the binning method, especially in the deep fields.

In \figref{fig:deep_panel}, we present real mosaics from the north deep field with $\pixsizeDeep$~pixels.
\figenv{
    \includegraphics[width=\textwidth]{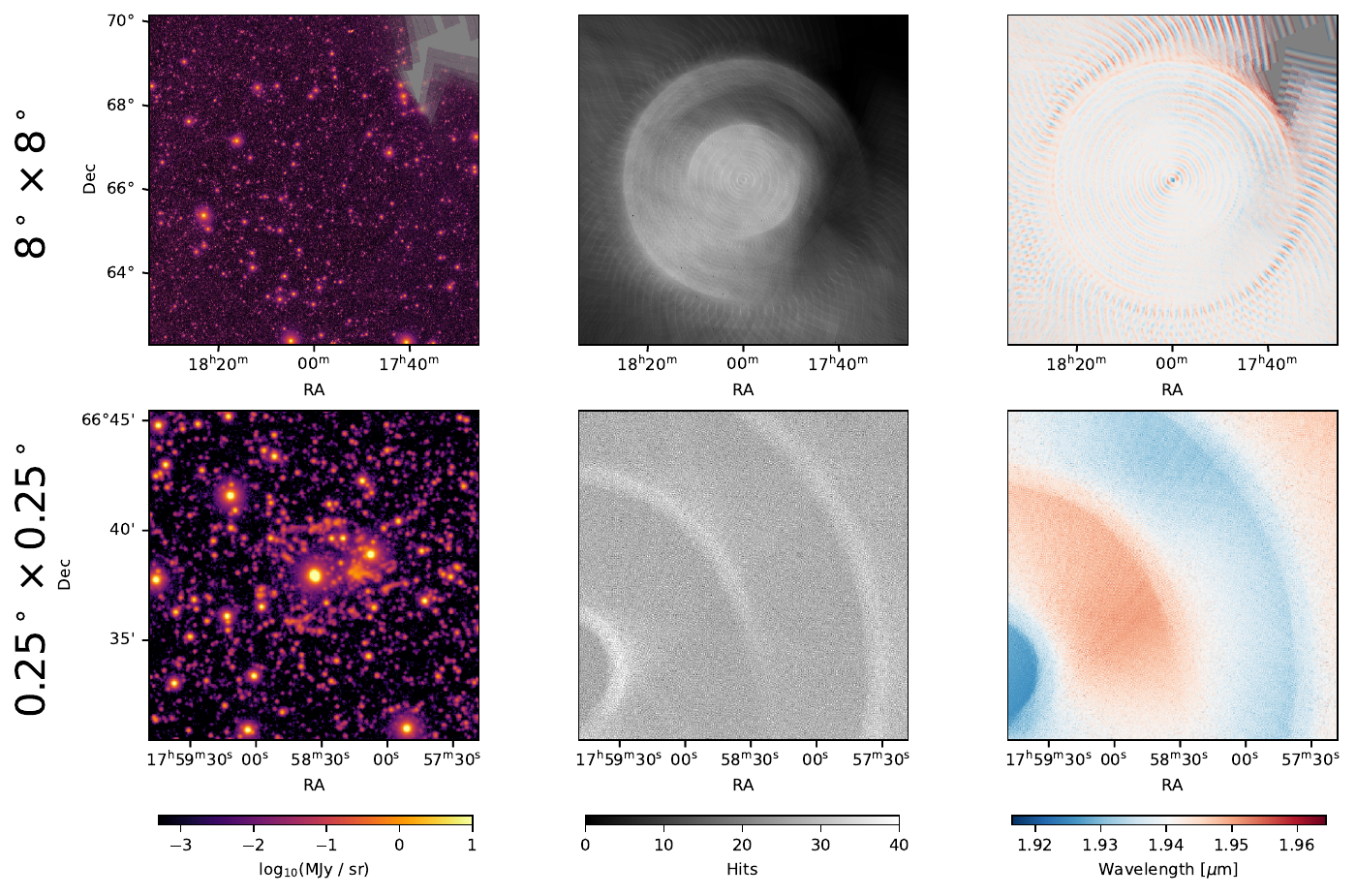}
    \caption{Deep-field mosaics for a channel centered on~$\wlDeep~\um$. The pixel size is~$\pixsizeDeep$. The maps have been filtered to remove zodiacal light and atmospheric emission~(\secref{sec:contourFilter}). With the large number of contributing exposures in the deep fields, relatively faint features can be extracted; notice that the color scale~(in the leftmost column) spans more than 4~orders of magnitude. The bright stars appear saturated only due to the choice of color scale. (\emph{Top row})~A relatively large area~($\dimsDeep$) centered on the north ecliptic pole. (\emph{Bottom row})~A smaller area~($\dimsDeepZoom$) containing the \catseye, which features an extended halo of~\Hmol.}
    \label{fig:deep_panel}
}
In the top row, we present an area of~$\dimsDeep$. In the bottom row, we zoom in to~$\dimsDeepZoom$ centered on the \catseye~(\catseyeCat), which is near the north ecliptic pole and serves as a useful target for map-making development. The \spherex\ telescope is unable to resolve the central feature for which the \catseye\ is named, but we can instead map the extended emission that stretches to a diameter of roughly~$\diamCatsEye$. We have chosen a spectral channel centered on~$\wlDeep~\um$ and covering the 1-0~S(3) transition of~\Hmol~($1.958~\um$), which produces an extended halo at a radius of \rExtHalo\ from the center. Image-level filtering~(\secref{sec:image_proc}) has been applied, and the filter residuals have been averaged down as a result of the repeated observations in the deep field. For filtering, an object mask has been applied to the entire extended halo of the \catseye\ in order to prevent distortions to its internal structure.

For \figref{fig:deep_panel}, we have incorporated L2~images through \fullskydate, which is approximately 7~months after the beginning of science observations. In the top row, there is a small region of missing coverage, which will be filled in by the end of the first year. At the boundaries of the coverage gap, the changing detector orientations can be seen most clearly. The hits maps~(middle column) are finer-resolution versions of those presented in \figref{fig:hits_panel}. Notice that the deep-field coverage is tiered with the most sensitive regions in the center.

In addition to sensitivity, the deep fields offer a suppression of the LVF gradients~(\secref{sec:spectral_images}). Compare the wavelength maps of the \heartsoul~(\figref{fig:coadd_panel}) to those of the north deep field~(\figref{fig:deep_panel}). For the latter, especially near the center, the wavelengths have averaged to be much closer to the center of the spectral channel. On the periphery of the deep field, where the hits are lower, we find that the LVF~gradients persist at a stronger level.

\subsection{Simulated Maps \label{sec:simmaps}}

Now that we have introduced the main map-making procedures, we return to the design goal of linearity~(\secref{sec:linearity}) and its value in studying individual simulation components~(\secref{sec:skysim}). Our image-processing~(\secref{sec:image_proc}) and map-making operations~(\secs{\ref{sec:reprojection} and \ref{sec:coaddition}}) have all been linear, so we can decompose our maps into their main contributions.

In \figref{fig:sim_panel}, we present maps of the seven main simulation components~(\secref{sec:skysim}) and their sum. 
\figenv{
    \includegraphics[width=\textwidth]{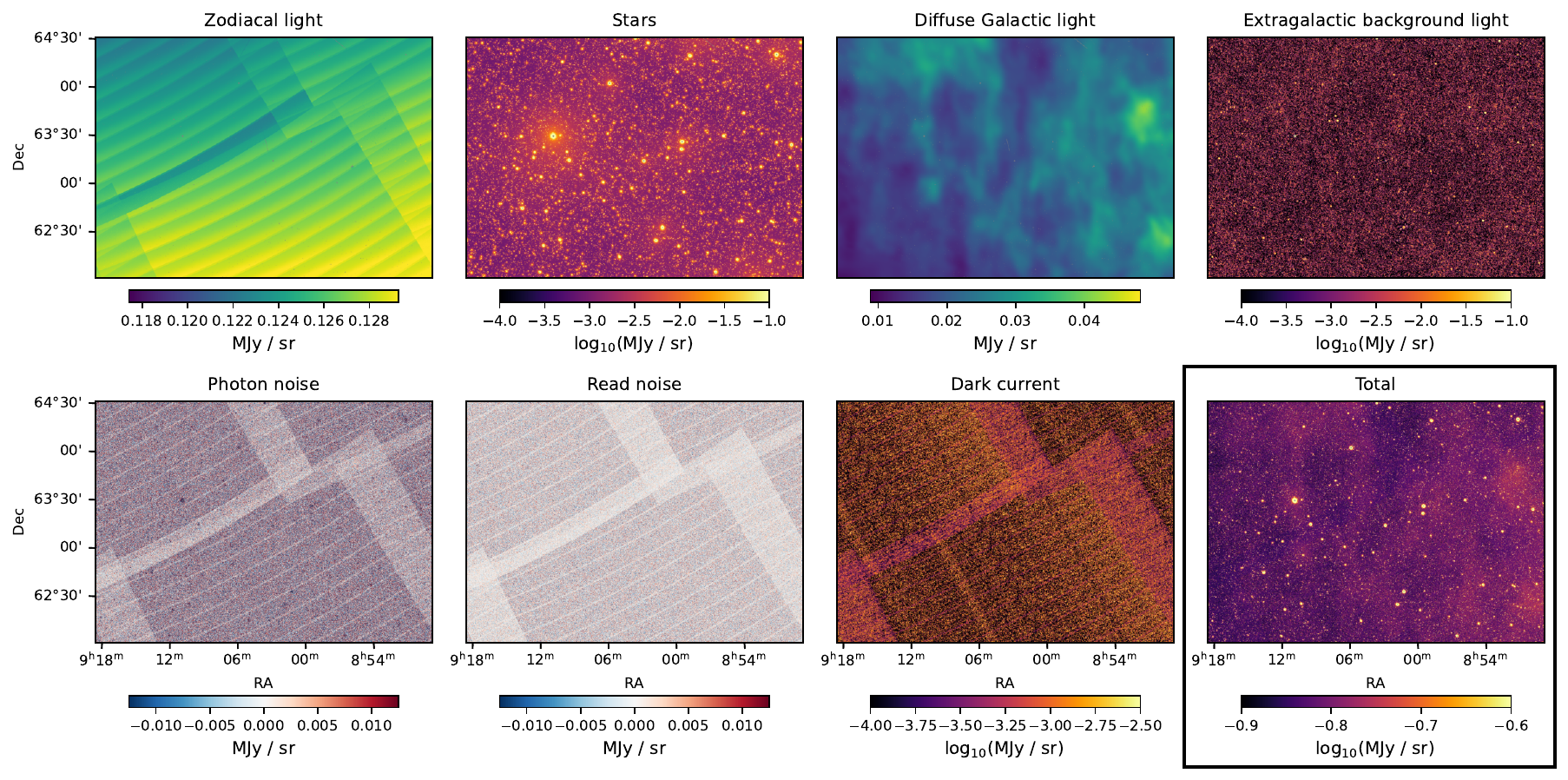}
    \caption{A region in Ursa Major with maps of individual simulation components and their sum; these have been evaluated for a spectral channel centered on~$\wlSim~\um$. The color schemes vary in order to best illustrate the features of the individual signals. Notice how the scan pattern manifests in several of the components. The contributing images were generated with the \skysim\ described in \cite{Crill2025}. Real flag masks~(\secref{sec:pixel_masking}) have been applied and are most noticeable at the centers of bright stars.}
    \label{fig:sim_panel}
}
These components correspond to the simulated images that appear in Fig.~5 of \cite{Crill2025}. For this example, we have chosen a region in Ursa Major that avoids any resolved objects of significant brightness; the \skysim\ is mainly designed for these types of regions. There is no image processing other than pixel-flag masking based on the real data. For this demonstration, we have selected a channel centered on~$\wlSim~\um$, which is in a wavelength range where both zodiacal light and DGL are significant.

We have deliberately chosen a region that is outside of the deep fields, so we can better illustrate the patterns produced by the observation-dependent components. The most significant of these is zodiacal light, which appears in the maps with LVF~gradients due to its~SED; the brightness variations are due to the position of the \spherex\ satellite within the zodiacal cloud at the time of observation. The zodiacal light varies by only about~$10\%$, but it appears in our maps with discrete discontinuities at a level that would overwhelm fainter features~(cf.~\figref{fig:deep_panel}). In regions of overlapping coverage, both photon noise and read noise tend to average down; the survey pattern is clearly visible in the noise components in \figref{fig:sim_panel}. To estimate map-level noise variances, an ensemble of noise realizations can be propagated, and summary statistics can be recorded. Photon noise is stronger than read noise and has been simulated from a combination of the four optical components. On large scales, the photon noise is dominated by zodiacal light; on small scales, by bright stars. Dark current averages in a more complicated manner that depends on the uniqueness of each pointing. In the L2~images~(\secref{sec:L2}), an in-flight estimate of dark current has been subtracted; since the estimate is approximate, it is occasionally helpful to restore the dark current and monitor it in simulation as in \figref{fig:sim_panel}.

For the example region of \figref{fig:sim_panel}, the small-scale features are dominated by starlight, while the large-scale features are dominated by DGL. Zodiacal light is brighter than DGL but much smoother. The EBL, which is here generated from a reference catalog of real~(rather than synthetic) galaxies, is dimmer but more statistically isotropic. In the bottom right, we present the sum of the seven simulation components.

The simulated maps are valuable mainly in understanding how known inputs are propagated through the map-making pipeline. For many purposes, the simulations need to be similar but not identical to the real data. At the time of writing, the \skysim\ produces starlight that is dimmer than in the real data and DGL that is brighter. These discrepancies will be reduced over time. It is not our aim to exactly imitate the real data, so we omit a comparison in \figref{fig:sim_panel} to the corresponding real mosaic.

\subsection{Map Uncertainties}

The notion of uncertainty depends strongly on the particular science goal, because each investigation has a different set of requirements and concerns. Faint galaxies, for example, could be considered to be part of a target signal like the extragalactic background light~(EBL), or they could be considered to be background fluctuations in the study of closer objects; this is the case in the spectroscopy of the \catseye\ in \secref{sec:spectroscopy}, for which background galaxies contribute to the estimated uncertainties. For many signals, \spherex\ is not limited by noise but instead by other effects such as unflagged transients, sparse spectral sampling or astrophysical foregrounds and backgrounds. We cannot define a unique pixel-level variance, and we instead describe tools that may be helpful in estimating uncertainties in a variety of cases.

Noise uncertainties can be estimated with the aid of the \skysim~(\secref{sec:skysim}), which produces images that can be propagated through the map maker. A single noise realization is presented in \figref{fig:sim_panel}; with multiple realizations, the ensemble statistics can be determined. Photon noise dominates, and it is predicted analytically from the expected sky signal. Read noise is estimated from non-optical, pre-launch testing.

It is important to note, however, that noise is only one source of variance and often subdominant. Although we provide an example of how to incorporate a noise model~(\secref{sec:simmaps} and \figref{fig:sim_panel}), we do not present this approach as generically reliable. The real data contain small- and large-scale transient signals, e.g., cosmic rays, satellite streaks, airglow and zodiacal light, which is effectively time-variable from the perspective of \spherex\ observations. There are also instrumental systematics such as PSF~variation, which can be relevant when the target object is small, and LVF~gradients, which can be relevant when the target displays strong emission lines. In these cases, variance could be estimated by splitting the dataset by observation period.  This type of approach is illustrated in \secref{sec:selfconsistency}, where we present a map difference as a way of estimating variance that includes noise, systematics and time-variable foregrounds.

In some cases, map-level uncertainties may be superfluous, because the relevant variances can be estimated from other elements of the analysis. This is the case in \cite{Hora2026}, for which uncertainties were estimated for aperture photometry, and this was done by characterizing the background scatter in the vicinity of each target region. A similar approach is taken for the spatially-resolved spectroscopy of the \catseye\ that we present in \secref{sec:spectroscopy}. In \cite{Murgia2026}, \spherex\ maps were correlated with each other and with an external dust tracer, and uncertainties were estimated for the correlation coefficients but not for any of the maps. Yet another approach is taken in the full-sky cross-correlations of \secref{sec:fullskycorr}, where variances are estimated by dividing the sky into bins of Galactic longitude.

\section{Spectral Data Cubes \label{sec:datacubes}}

Thus far, we have considered only a single spectral channel for each example region, but much of the scientific power of \spherex\ lies in its broad wavelength coverage. We now describe the construction of spectral data cubes, i.e., collections of maps corresponding to multiple different wavelengths. Nominally, \spherex\ divides each detector array into \nsubchs~channels; with six arrays, this produces \nchs~maps.

\subsection{Mosaic Ensembles \label{sec:mosaicens}}

As an example of a spectral data cube, we consider an ensemble of mosaics of the \catseye. In \figref{fig:deep_panel}, we presented a mosaic of the \catseye\ for a single channel in an area of~$\dimsDeepZoom$. The same map-making operations can be applied to all of the spectral channels, and we present them together in \figref{fig:datacube}.
\figenv{
    \includegraphics[width=0.91\textwidth]{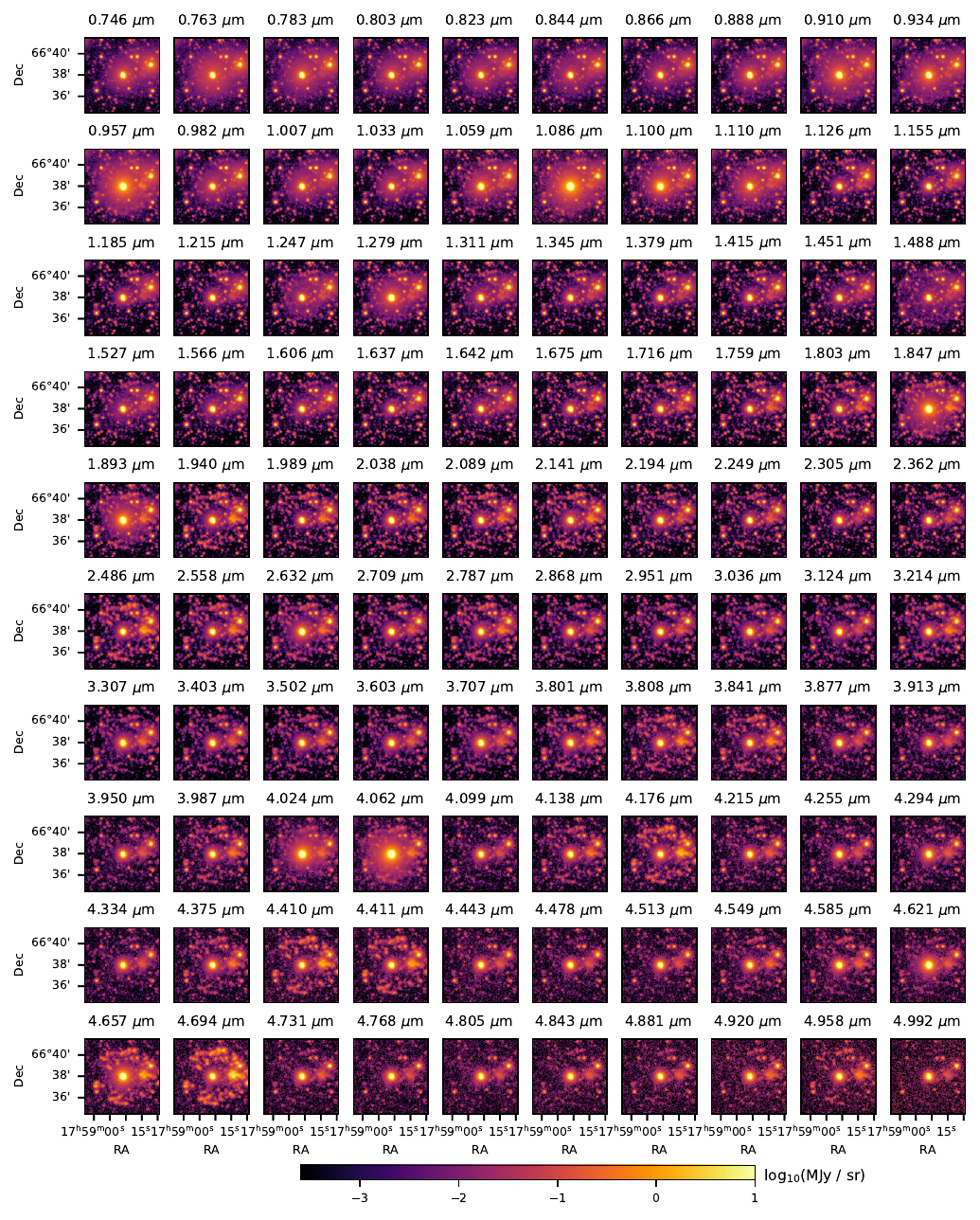}
    \caption{Spectral data cube of the \catseye\ with \nChsCatsEye~channels. The processing is identical to that of the maps presented in \figref{fig:deep_panel}, but we have cropped to~$\dimsCatsEye$ in order to focus on the nebular features. Notice the spectral-spatial variation that is captured by this mosaic ensemble. These maps cover more than 4~orders of magnitude in intensity. The central star saturates the color scale but not the \spherex\ detectors; we restricted the color maximum to better emphasize the fainter features of the extended halo.}
    \label{fig:datacube}
}
We have removed the last channel of array~3 and the first channel of array~4, because they are non-negligibly affected by gain uncertainties related to the DBS in the \spherex\ optical system~\citep{Korngut2026}. In \figref{fig:datacube}, then, the number of maps is~\nChsCatsEye. 

The maps in \figref{fig:datacube} are presented in order of ascending wavelength. Because of small wavelength overlaps between neighboring arrays, the first channel of array~2 appears ahead of the last channel of array~1. Similar reorderings occur at the other array boundaries. 

Because image-level filtering has been applied, the zodiacal-light residuals are, on average, zero. At the longer wavelengths, zodiacal light is brighter~(due to the Wien tail of the thermal component), so the residuals and the associated photon noise tend to be stronger. Additionally, the longer-wavelength channels have fractionally smaller passbands~(larger spectral resolution) and are, therefore, noisier. In \figref{fig:datacube}, the color scale is held fixed for all channels and extends to small enough intensities to see the noise increase at long wavelengths. 

The image-level filtering~(\secref{sec:contourFilter}) has largely removed the helium airglow~(\secref{sec:spectral_images}) that dominates near~$\wlHe~\um$ in the unprocessed images. In that wavelength range, the filtered mosaics reveal an extended, azimuthally symmetric brightening that is likely due to \HeI\ in the \catseye\ itself. Although the filter removes airglow on the average, the photon noise remains. Upon close examination, the $1.086$-$\um$ mosaic, which is closest to the \HeI\ transition, displays an elevated background variance compared with the mosaics of nearby channels. Other atmospheric effects are weaker but still mitigated by the contour filter.

Across the \spherex\ wavelength range~($\wlmin$-$\wlmax~\um$), the \catseye\ displays significant variation in its inner and outer halos, which extend to approximately~$\diamCatsEye$ in diameter. For example, the inner halo~(within $\sim 1\arcmin$) brightens at wavelengths associated with the \Paa~($\wlPaa~\um$), \Bra~($\wlBra~\um$) and \Pfb~($\wlPfb~\um$) transitions of~\HI. We find similar brightenings for \HeI~($\wlHe~\um$) and [\SIII]~($\wlSIIIdim$ and $\wlSIII~\um$). A subtler inner-halo enhancement can also be visually identified for [\ArIII]~($\wlArIII~\um$). The outer halo~(\rExtHalo) brightens at wavelengths that are primarily associated with hot~\Hmol. The strongest \Hmol~enhancement corresponds to the \HmolTransEx~transition at~$\wlHmolEx~\um$, but many other transitions produce similar features. In \secref{sec:spectroscopy}, we identify a larger but non-exhaustive set of spectral features.

The mosaic ensemble in \figref{fig:datacube} is meant to serve as a small demonstration of the capabilities of spectral map making with \spherex. Similar data cubes can be constructed for any region of the sky, though the processing choices may need to be altered depending on the signals of interest. The \catseye\ is arguably a special case due to its location in one of \spherex's deep fields. Other regions have less coverage and, therefore, less averaging. Noise and filter residuals will be larger, and LVF~gradients will be stronger. In \figref{fig:datacube}, we have used the nominal choice of $\nsubchs$~channels per array~(with the two DBS-adjacent channels removed), but other targets may be better studied with channels that are wider or narrower~(\secref{sec:chdef}).

\subsection{Spatially-resolved Spectroscopy \label{sec:spectroscopy}}

To illustrate the spectral content of \spherex\ data cubes, we perform spatially-resolved spectroscopy on the mosaic ensemble of the \catseye~(\figref{fig:datacube}). The results are presented in \figref{fig:aperturephotometry}, where we have defined three effective apertures for photometry: a small region in the core~(red), an annulus in the inner halo~(green) and a small region in the outer halo~(blue).
\figenv{
    \gridline{
        \fig{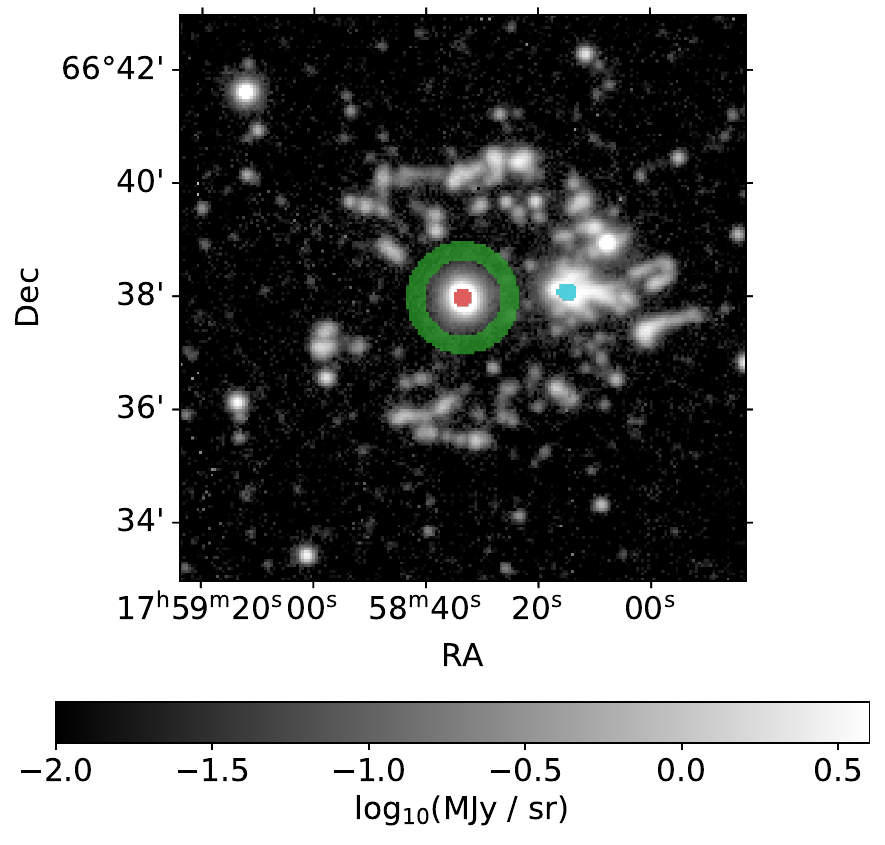}{0.37\textwidth}{}
        \fig{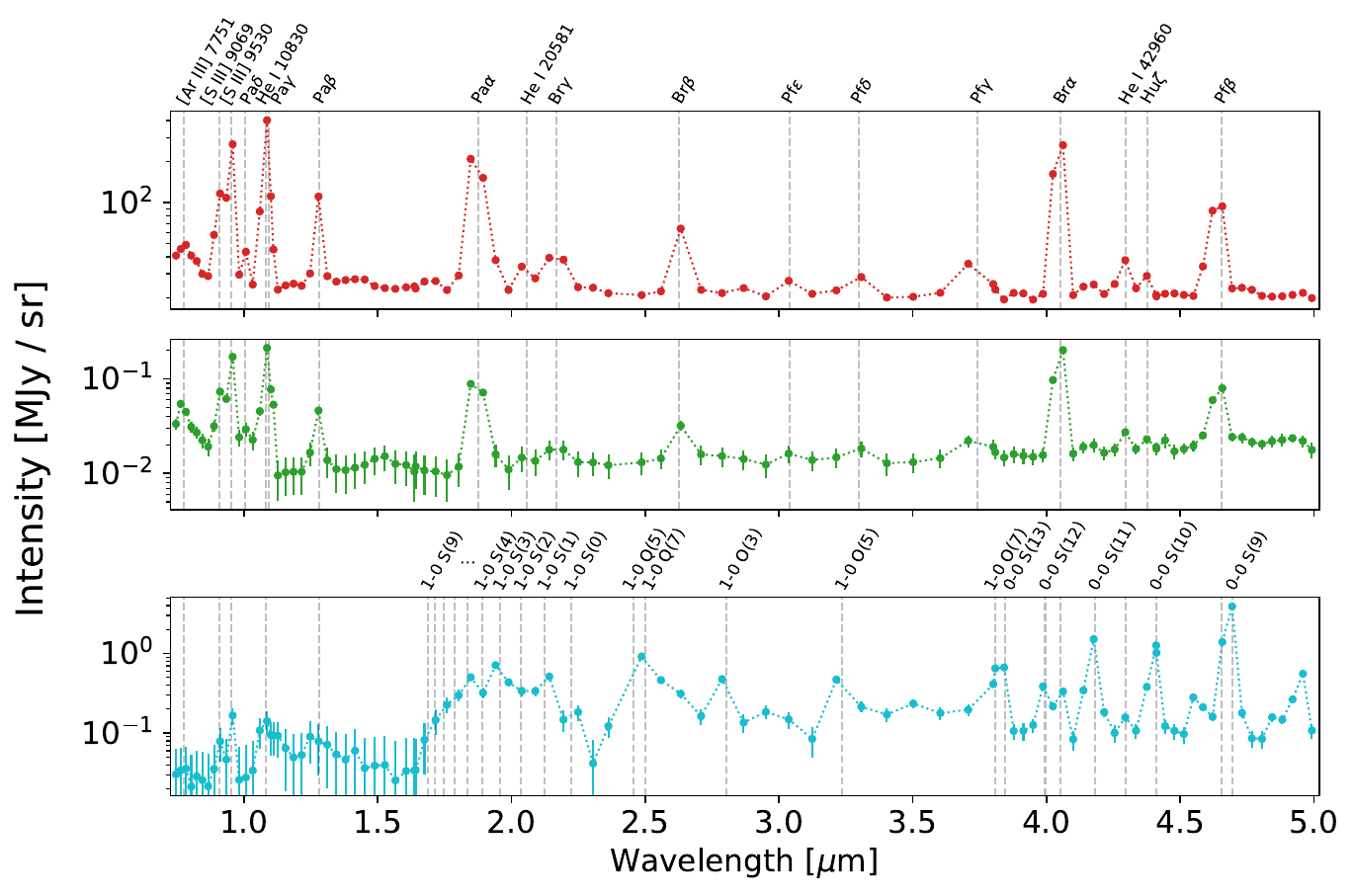}{0.61\textwidth}{}
    }
    \caption{Spatially-resolved spectroscopy of the \catseye. (\emph{Left})~Three apertures are defined with an example mosaic~(centered on $4.694~\um$) in the background. We define an aperture for the core~(red), the inner halo~(green) and the outer halo~(blue). (\emph{Right})~The corresponding spectra are assembled from the mosaic ensemble~(\figref{fig:datacube}). We have labeled the most prominent spectral features with probable identifications. In the bottom panel, the emission features near~$4.55$ and~$4.96~\um$ are possibly due to an accumulation of lines associated with~carbon monoxide~(\CO).}
    \label{fig:aperturephotometry}
}
For each aperture, we calculate the mean value, and then we collect the results from all of the channel maps in order to form a spectrum. For the inner and outer halo, we additionally apply a source mask in order to suppress the influence of unrelated stars and galaxies. We also characterize the background by performing the same operations at random locations outside of the extended halo of the \catseye. We subtract the background mean, and we report the standard deviation as the photometric uncertainty.

There are a number of nonidealities in the spectra of \figref{fig:aperturephotometry}. We have deconvolved neither the PSFs nor the passbands, so the relative line strengths may be misleading. In particular, recall that arrays~5 and~6~($\sim 3.8$-$5.0~\um$) have higher spectral resolution, so emission lines are less diluted. At the same time, the PSF generally increases with wavelength, and that will weaken point-source intensities and also leak some contamination outside of the source masks. Since the \catseye\ is near the center of the north deep field, the LVF~gradients have averaged down significantly~(\figref{fig:deep_panel}), but there is still some residual wavelength variation. It is always possible to revert to the L2~images and use the exact, unbinned wavelengths. The data cube is essentially a compression of the L2~images, and it may be simpler to handle but is subject to certain approximations. In \cite{Hui2026}, the \spherex\ spectral calibration is tested by comparison with the strongest emission lines in the core of the \catseye; the test uses a spectrum that is based directly on the L2~images and is comparable to the red curve in \figref{fig:aperturephotometry} but with finer wavelength sampling. Recall, however, that \spherex's spectral resolution is limited mainly by the LVF~passbands~(\secref{sec:spectral_images}) rather than the wavelength sampling.

Many prominent spectral features are evident in \figref{fig:aperturephotometry}, and \spherex\ is sensitive to these over more than 4~orders of magnitude in intensity. With \spherex's spectral resolution, emission lines are unresolved and sometimes overlapping. In the core and inner halo, we find strong lines associated with partially ionized gas; we have labeled some of the transitions for~\HI, \HeI, \SIII\ and~\ArIII. In the outer halo, the spectrum is dominated by lines associated with hot~\Hmol\ and possibly carbon monoxide~(\CO). Some of these features have been identified in previous studies~\citep[e.g.,][]{Hora1999,Hyung2000,Hora2004,Klaas2006}. Our spatially-resolved spectroscopy may enable a novel investigation of the extended halo of the \catseye, but we defer a detailed analysis to future work.

\subsection{Continuum Subtraction \label{sec:continuum_subtraction}}

The \spherex\ data cubes enable map-space isolation of spectral features. With the large number of wavelengths, we can fit and subtract a local continuum contribution. 

In \figref{fig:contsub}, we present an example of continuum subtraction for the \tri~(\tricat) in order to reveal PAH~emission at~$\wlPAH~\um$.
\figenv{
    \plotone{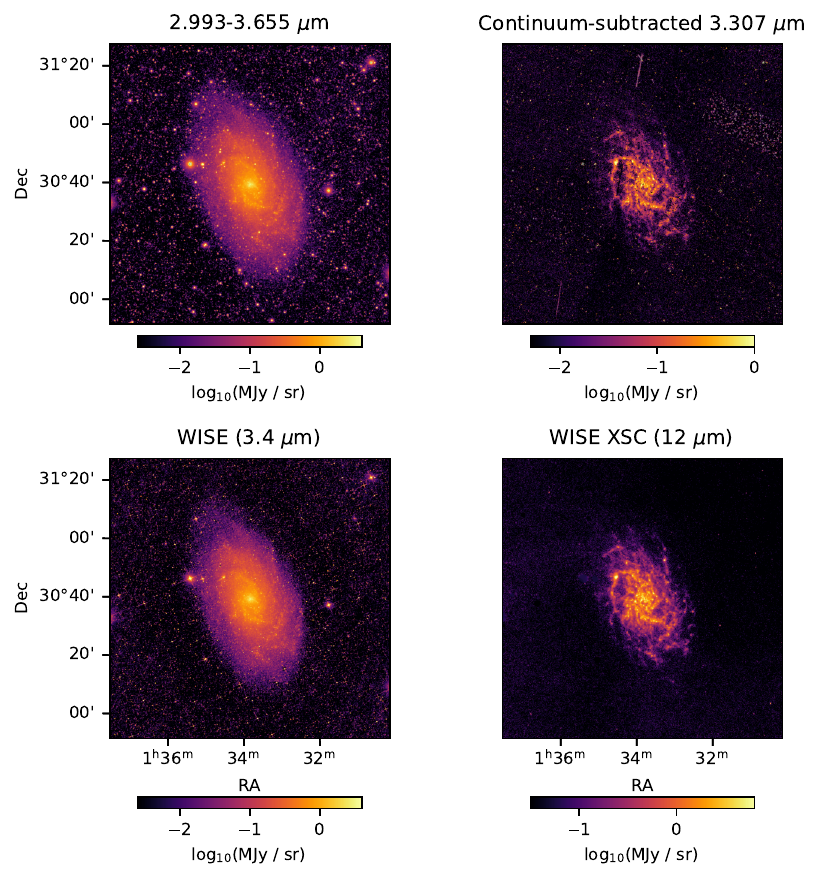}
    \caption{Continuum subtraction to reveal PAH emission in the \tri~(\tricat) and a comparison to similar maps from \wise. (\emph{Top left})~Broadband map made from seven nominal channels covering approximately the same wavelength range as \wise~\Wone~(centered on~$\wlWone~\um$). (\emph{Bottom left})~The corresponding map from \wise\ with the same color scale for direct comparison. (\emph{Top right})~Continuum-subtracted map revealing $\wlPAH$-$\um$~PAH emission. (\emph{Bottom right})~The $\wlWthree$-$\um$~(W3) map from the \wxscFull; this band is sensitive to a different set of PAH emission features, which produce similar spatial structures. The color scale is different but spans the same logarithmic distance.}
    \label{fig:contsub}
}
The results are based on a mosaic ensemble that is similar to that of \figref{fig:datacube}. We have applied image-level filtering~(\secref{sec:contourFilter}) with an object mask~(\secref{sec:pixel_masking}) to preserve the structure of~\tricat~(cf.~\figref{fig:filter_panel}, which features~\pinwheelcat). The maps are made with the nominal channel definitions~(\nsubchs~per array). The PAH~emission is represented by the channel that is most closely centered on~$\wlPAH~\um$. The PAH~emission profile is quite broad~\citep[e.g.,][]{Lai2020}, so we must ensure that the continuum is estimated from channels that are sufficiently removed from the peak. In this case, we use~$\wlMinContsub$ and~$\wlMaxContsub~\um$ to fit and subtract a linear continuum. This operation reveals the spatial structures associated with $\wlPAH$-$\um$~PAH emission~(in the upper-right panel of \figref{fig:contsub}).

Since we approximated the continuum to be linear, the subtracted map can contain residuals, especially at the locations of bright stars. The results may be improved by fitting the continuum with more degrees of freedom and by using more of the image-level wavelength information in order to fit the PAH~profile~\citep{Hora2026}.

We compare our results to maps produced from the Wide-field Infrared Survey Explorer~\citep[\wise,][]{Wright2010}. The \wise\ bands were much broader than \spherex's, so we combine seven nominal \spherex\ channels~(\secref{sec:chdef}) in order to approximate the \Wone~band, which is centered on~$\wlWone~\um$. Because the \wise\ telescope aperture is twice as large as that of \spherex, stars appear sharper in the \wise\ map. Due to differences in the optical designs, bright stars produce diffraction spikes in \wise\ but not in \spherex. Although the \Wone~map contains the $\wlPAH$-$\um$~PAH emission feature, it is difficult to isolate it from the continuum. We instead compare to the $\wlWthree$-$\um$~(\Wthree)~observations, which contain contributions from a different set of PAH~emission features~\citep[e.g.,][]{Tielens2008}. We expect the PAH~morphology to be similar but not identical. Because our continuum subtraction removes most of the signals from point sources, we compare to the \tricat~map from the \wxscFull~\citep[\wxsc,][]{Jarrett2019}, which has removed point sources as well. A similar comparison could have been made to the $\wlWthree$-$\um$~dust maps of \cite{Meisner2014}, which also used~\Wthree. We find, in the lower-right panel of \figref{fig:contsub}, that the PAH~morphologies are broadly consistent between~$\wlPAH$ and~$\wlWthree~\um$. In the dimmer regions, there may be a deficit at~$\wlPAH$ as compared to~$\wlWthree~\um$. We note that this type of morphological analysis can be performed on any object that is resolved by \spherex, but we defer such comparative studies to future work.

Similar continuum subtraction can be performed on any region of the sky. See \fnref{fn:allsky} for full-sky maps of PAHs and~\HII. See \cite{Murgia2026} for PAH and \Bra\ maps of the \orion~(\orioncat) and the Galactic plane. In \cite{Hora2026}, similar techniques are used to extract features representing both emission~(e.g., from PAHs, \HII~regions and~\Hmol) and absorption~(e.g., from \water\ and \COtwo~ice).

\subsection{Filtering Considerations \label{sec:filter_considerations}}

Image-level filtering~(\secref{sec:contourFilter}) is effective in removing zodiacal and atmospheric foregrounds, but it can also remove part of the target signal. For small objects like the \catseye~(\figs{\ref{fig:deep_panel} and~\ref{fig:datacube}}) and \tricat~(\figref{fig:contsub}), masks can be applied at the image level to ensure that the filter is influenced only by the surrounding sky~(cf.~\figref{fig:filter_panel}). When the target is larger than the \spherex\ field of view, e.g., for the \heartsoul~(\figref{fig:coadd_panel}), then it is impossible to mask at the image level~(without removing all of the information). Without a mask, the filter can still be applied, but it will remove information and distort the target object.

In general, even for compact objects that could be masked at the image level, it is preferable to avoid filtering when possible. Even if the object is masked, the background filtering may impart offsets that vary from image to image. The filter is mostly easily avoided when the target signal is bright relative to zodiacal light and atmospheric emission. An example is the \orion, which is mapped without filtering in \cite{Murgia2026}. The image-level filter could also be avoided if the foregrounds will be handled at the level of the mosaics, e.g., through a deprojection of simulated zodiacal light~(\secs{\ref{sec:deproj} and~\ref{sec:fullskydeproj}}).

Yet another case is when the target is a prominent spectral feature that will be isolated by continuum subtraction~(\secref{sec:continuum_subtraction}). Although zodiacal light is expected to be spectrally smooth over most of the \spherex\ wavelength range, it appears in \spherex\ maps with an amplitude that depends on time. The \spherex\ spatial coverage depends on both time and wavelength, so spectral smoothness can be disrupted in unfiltered maps. For a small region, however, most observations occur within a relatively small time period~(within each full-sky survey as described in \secref{sec:scan}), so the time dependence may be relatively weak. In this case, mosaic-level continuum subtraction may suppress zodiacal light relative to sharper spectral features like hydrogen recombination lines or $\wlPAH$-$\um$~PAH emission.

In some cases, however, image-level filtering is crucial. In \figref{fig:filtcoadd_panel}, we consider one of the most challenging examples of foreground contamination. 
\figenv{
    \plotone{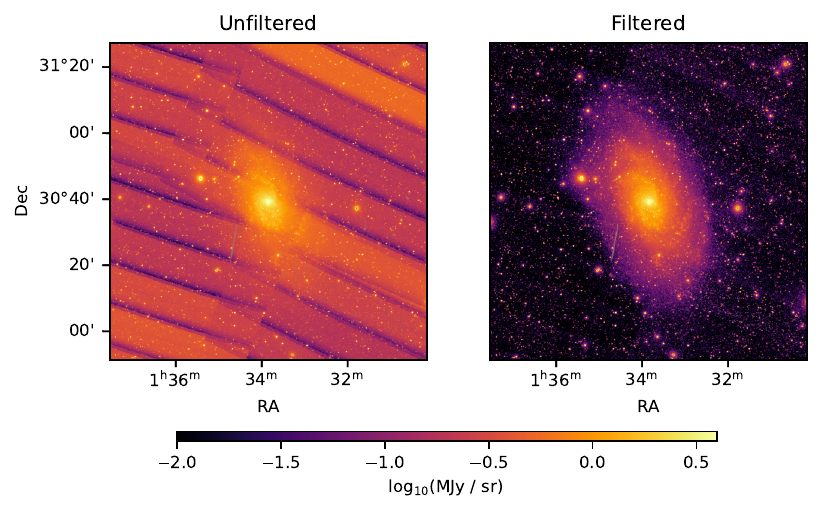}
    \caption{Illustration of the benefits of image-level filtering for the especially challenging case of helium airglow at~$\wlHe~\um$. (\emph{Left})~The mosaic without filtering. In order to emphasize the non-uniform nature of the airglow contamination, we have subtracted an offset~($0.3~\MJysr$). (\emph{Right})~The mosaic with filtering. Some faint residuals are visible in the background~(e.g., in the upper right). Negative values, which are hidden by the logarithmic color scale, have been separately verified to be small compared with the features of~\tricat.}
    \label{fig:filtcoadd_panel}
}
We map~\tricat\ at~$\wlHe~\um$, which is the wavelength of the helium airglow that appears as a bright LVF~band in all array-1 images~(e.g.,~\figs{\ref{fig:exp_panel} and \ref{fig:filter_panel}}). The airglow is strong and difficult to model~\citep{airglow2026}. Without image-level filtering, mosaics are severely contaminated with striping features that are determined by the survey pattern and the temporal-spatial variation in the airglow amplitude. This is illustrated in the left panel of \figref{fig:filtcoadd_panel}, where \tricat~is overwhelmed by the unmitigated airglow except in the bright center. At the time of writing, the airglow cannot be modeled to a precision that would allow for a direct subtraction. If we instead apply image-level filtering~(as in, e.g., \figref{fig:filter_panel}) and construct a new mosaic, we are able to significantly suppress the airglow and reveal the entirety of~\tricat~(right panel of \figref{fig:filtcoadd_panel}). In the background, airglow residuals are faintly visible.

For intensity mapping, which is among the main-mission goals of \spherex~\citep{Bock2026}, filtering is necessary in order to achieve the requisite sensitivity, but maps will generally be larger than the \spherex\ field of view~(e.g.,~the top row of \figref{fig:deep_panel}). In order to estimate EBL~fluctuations on large scales, it is important to characterize the transfer function of the filtering process, i.e., the suppression of features as a function of angular scale. With the \skysim~(\secref{sec:skysim}), we can propagate known input signals through all image-processing and map-making steps~(\secref{sec:simmaps}) and monitor scale-dependent distortions. The transfer function can then be corrected when estimating summary statistics like power spectra.

\subsection{Spectral Interpolation \label{sec:specInterp}}

We now confront the issue of the LVF~gradients that cause an entanglement of spatial and spectral features. In regions with bright emission lines, this effect is especially pronounced and produces unphysical distortions and discontinuities. By defining channels that are narrower than nominal~(\secref{sec:chdef}), we can spectrally interpolate to a target wavelength and suppress the LVF-induced artifacts.

We first consider the conditions under which the LVF~gradients are most problematic. In regions with many repeated visits, the gradients tend to average down as illustrated in, e.g., \figref{fig:deep_panel} for the north deep field. Most of the sky, however, has coverage that is relatively shallow, and the LVF~gradients persist at nearly full strength as in, e.g., \figref{fig:coadd_panel} for the \heartsoul. The  gradients are tolerable when there is relatively little spectral variation near the target wavelength; for \figref{fig:coadd_panel}, we deliberately chose a wavelength that satisfies this criterion. When observing emission lines or other sharp spectral features, the wavelength variation can create unphysical scan patterns in the mosaics, because the observed line strength depends very sensitively on the central wavelength of each detector pixel~(exemplified by the horizontal stripes in \figref{fig:exp_panel}). In such cases, we can no longer ignore the small wavelength spread within a spectral channel.

In \figref{fig:interp_panel}, we consider an example of a prominent emission line that creates unphysical map-level gradients~\citep[referred to as ``banding'' in][]{Hora2026}. 
\figenv{
    \includegraphics[width=\textwidth]{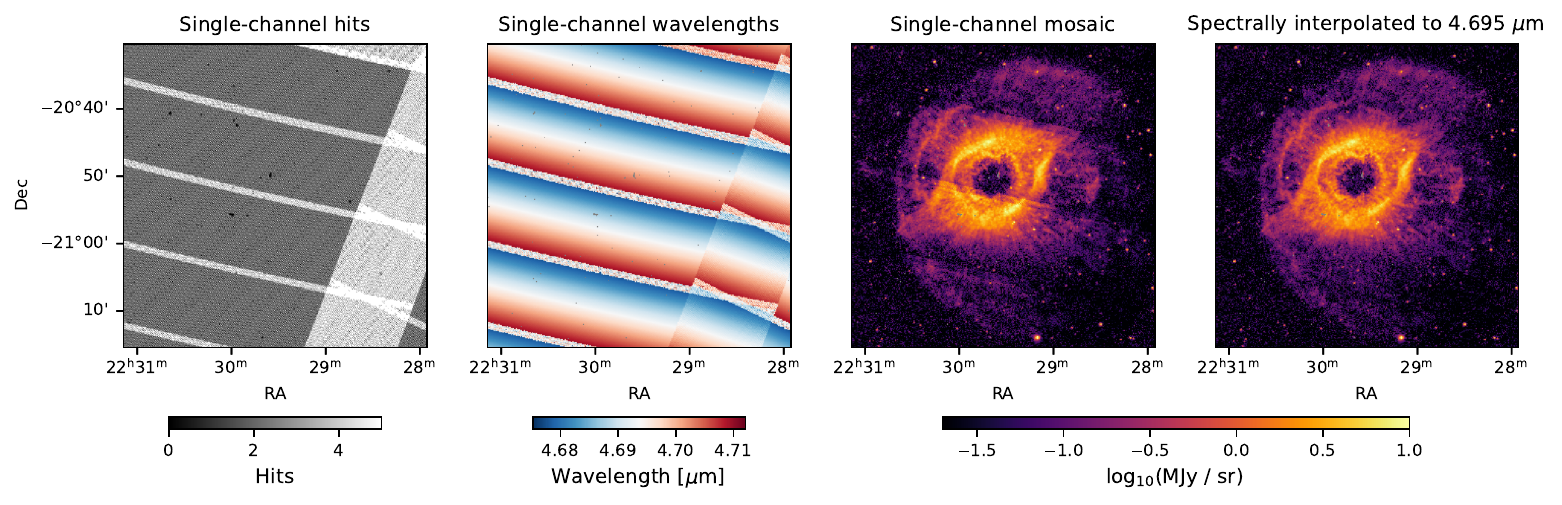}
    \caption{Illustration of spectral interpolation with the \helix. The left three panels use a nominal spectral channel centered on~$\wlChInterp~\um$, which is near the \HmolTransEx~transition of~\Hmol~($\wlHmolEx~\um$). The coverage is restricted to \fullskydate\ in order to emphasize the phenomenon that spectral interpolation addresses. (\emph{Left})~Hits map for the nominal channel. (\emph{Second from left})~Detector wavelengths that contribute to each mosaic pixel. (\emph{Third from left})~Nominal channel map showing LVF-induced gradients and discontinuities. (\emph{Right})~Spectrally interpolated map evaluated at~$\wlHmolEx~\um$; the most severe features have been significantly suppressed.}
    \label{fig:interp_panel}
}
Our example considers the \helix~(\helixcat) near $\wlHmolEx~\um$, which is associated with the \HmolTransEx\ transition of \Hmol. If we make a map with the spectral channel that contains the \HmolTransEx\ transition, then the result suffers from sawtooth patterns induced by the LVF~gradients. In the leftmost panel, we provide the hits map, which shows that the coverage is relatively shallow~(as of \fullskydate). Correspondingly, in the panel that is second from left, we show the map of detector wavelengths that observed each map pixel. The wavelength spread is relatively small, but the \HmolTransEx\ transition is quite sharp, even when observed with the relatively broad \spherex\ passbands. In the third panel from left, we present the channel map, which displays discrete discontinuities that correspond to the heterogeneous wavelength coverage. As \spherex\ collects more observations, the wavelength coverage will tend to homogenize as in the centers of the deep fields~(\figref{fig:deep_panel}). For most sightlines, however, \spherex\ returns only twice per year~(\secref{sec:scan}), so the wavelength homogenization is relatively slow.

A similar problem was encountered and mitigated in \cite{Hora2026}, which focused mainly on ice absorption and PAH~emission in the Cygnus-X region. Corrections were applied at the image level and were based on the detector passbands as well as the expected profiles of the spectral features. With the corrected images, new mosaics were constructed with ``banding'' artifacts that were significantly suppressed. In this work, we describe an alternative technique that may be preferable when the spectral profiles are uncertain or when it is desirable to work with a smaller number of data products.

Our approach is illustrated in \figref{fig:narrowband_mosaics}, where we construct narrowband mosaics in order to spectrally interpolate to the target wavelength. 
\figenv{
    \includegraphics[width=\textwidth]{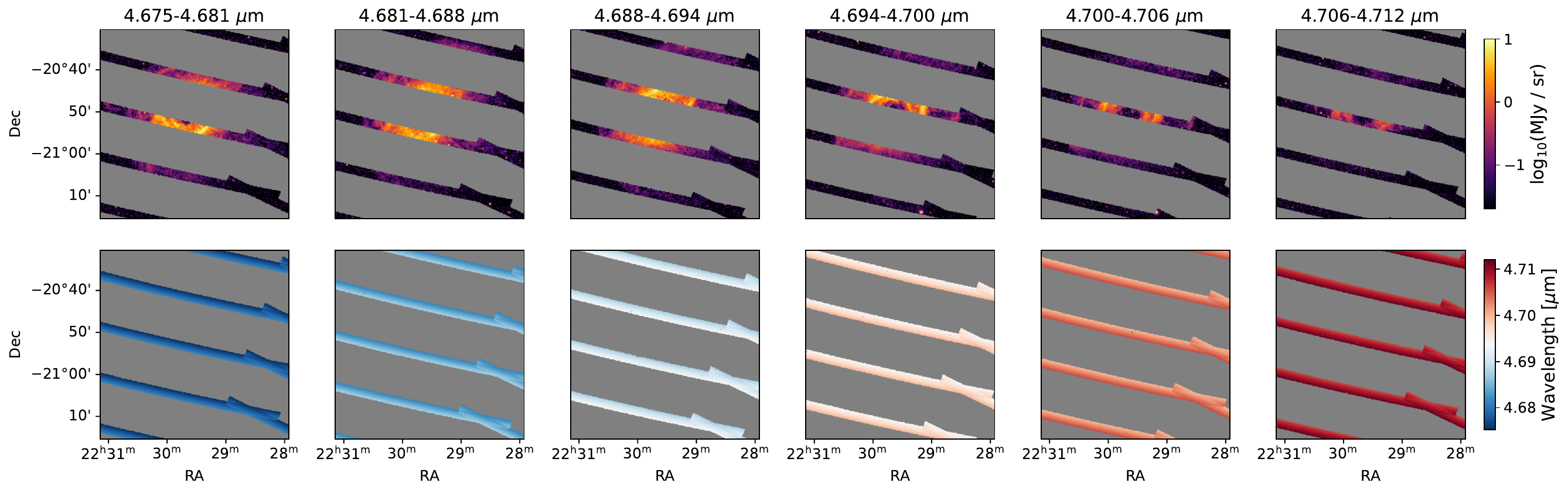}
    \caption{Some of the narrowband mosaics that can be used for spectral interpolation of the \helix. In this example, a nominal channel map~(third from left in \figref{fig:interp_panel}) has been divided into six parts, each of which covers a smaller range of wavelengths. This comes at the cost of spatial coverage per mosaic. (\emph{Top row})~Narrowband mosaics. (\emph{Bottom row})~Detector wavelengths that contribute to each narrowband mosaic.}
    \label{fig:narrowband_mosaics}
}
For this example, we have divided each nominal channel into six equal parts, so there are now \nsubchsInterp~channels per array instead of~\nsubchs~(cf.~\figref{fig:ChDef}). In \figref{fig:narrowband_mosaics}, we show how a nominal channel map~(third from left in \figref{fig:interp_panel}) is decomposed into six narrowband mosaics, each of which is spatially sparser but spectrally purer. There is still some wavelength variation within each of these narrowband mosaics, but they more closely approximate monochromaticity. With this new data cube of narrowband mosaics, we interpolate along each mosaic pixel to estimate the intensity at a target wavelength, which in this case is~$\wlHmolEx~\um$, the \HmolTransEx~transition of~\Hmol. The interpolation relies only on wavelengths that are relatively close to the target; typically, no more than 30~narrowband mosaics contribute, and the number is independent of the mosaic area. The interpolating function is a cubic spline that is forced to be monotonic between data points; this prevents unphysical oscillations but may also bias the results. We reiterate that this type of interpolation is most useful when the true spectral profiles are uncertain or when it is more convenient to work with a relatively small set of mosaics rather than a relatively large number of L2~images. The LVF~gradients will be suppressed, but additional validation is required in order to assess the accuracy of the interpolation.
For the \helix, our interpolated result is shown in \figref{fig:interp_panel} in the rightmost panel, where the sawtooth pattern has been significantly suppressed.

Spectral interpolation is most effective in removing the discrete discontinuities imparted by the LVF-based survey strategy. The interpolation is informed by a limited amount of spectral information, so scan-correlated features may persist at a lower level. As the \spherex\ survey continues, more spectral information will become available, and the interpolation will be better constrained. If one can model the spectral profiles of the strongest components, interpolation can be replaced by a fit to narrowband mosaics like those of \figref{fig:narrowband_mosaics}; the best-fit model can then be evaluated at the target wavelength.

\subsection{Deprojection \label{sec:deproj}}

Through a process we term ``deprojection'', we can fit and remove foregrounds, backgrounds or other confounding components. For example, in studying the~EBL, it may be desirable to remove starlight and~DGL. We can form templates for these components and then subtract scaled versions that minimize the variance over the mosaic pixels.

In the case of wide-area mapping, i.e., on scales larger than the \spherex\ field of view, it may be necessary to avoid image-level filtering in order to preserve the features of interest. This, however, leaves zodiacal light in the resulting mosaics. For bright features, the zodiacal-light contamination may be negligible and tolerable; for example, both \cite{Hora2026} and \cite{Murgia2026} consider regions near the Galactic plane, where zodiacal light is subdominant and can be left unfiltered. For fainter regions, however, the zodiacal light must be mitigated, and one approach is to implement deprojection with a template defined by the predictions of the \skysim~(\secref{sec:skysim} and \figref{fig:sim_panel}). In \secref{sec:fullskydeproj}, we will demonstrate this type of deprojection with maps of the full sky.

\section{Full-sky Maps \label{sec:fullskymaps}}

We now describe methods and considerations for mapping the full sky. The \spherex\ \st\ has released visualizations of full-sky maps with data through \fullskydate~(\fnref{fn:allsky}); these include maps that have been continuum subtracted~(\secref{sec:continuum_subtraction}) to reveal PAHs and~\HII. All-sky data cubes are planned to be released 6~months after the conclusion of the first year of science observations~\citep{Akeson2025}.

For full-sky maps, we prefer pixelizations that are significantly coarser than \spherex's detector pixels~($\pixsize$). The full survey would tile the sky into approximately $1.4 \times 10^{10}$~detector pixels. In the popular \healpix\ representation~\citep{Gorski2005}, the pixelization is controlled by the parameter~$\nside$, which is conventionally a power of~$2$. We can achieve $6\farcs4$~pixels with $\nside=32768$. Compare this to the \planck\ maps~\citep[e.g.,][]{Planck2018I}, which are typically released with~$\nside=2048$~(corresponding to a pixel side length of~$1\farcm7$). With modern computation, much larger pixel counts are now possible, but data processing is still significantly faster with products that are smaller. In our discussion of full-sky maps, we focus on scales much larger than \spherex's detector pixels. This will allow us to dramatically reduce the pixel count. We also aim for a convenient format for cross-correlations with other full-sky maps such as those sensitive to the cosmic infrared background~(CIB), thermal dust emission or 21-cm~\HI~emission~\citep[e.g.,][]{PlanckXLVIII2016,HI4PI2016}.

For full-sky maps, our reprojection algorithm is the same binning approach that we described in \secref{sec:reprojection}, but now the map grid is defined by \healpix. This method is used in \cite{Murgia2026}, which focuses its analysis on the Galactic plane but which also presents full-sky maps for a small number of spectral channels. Before reprojection, we smooth the images to a scale comparable to the relatively large \healpix\ pixels; this will be described in more detail in \secref{sec:smoothing}.

\subsection{Full-sky Data Cube \label{sec:fullsky_datacube}}

In \figref{fig:fullsky_datacube}, we present full-sky maps for \nChsFullSky~spectral channels with data through \fullskydate.
\figenv{
    \includegraphics[width=\textwidth]{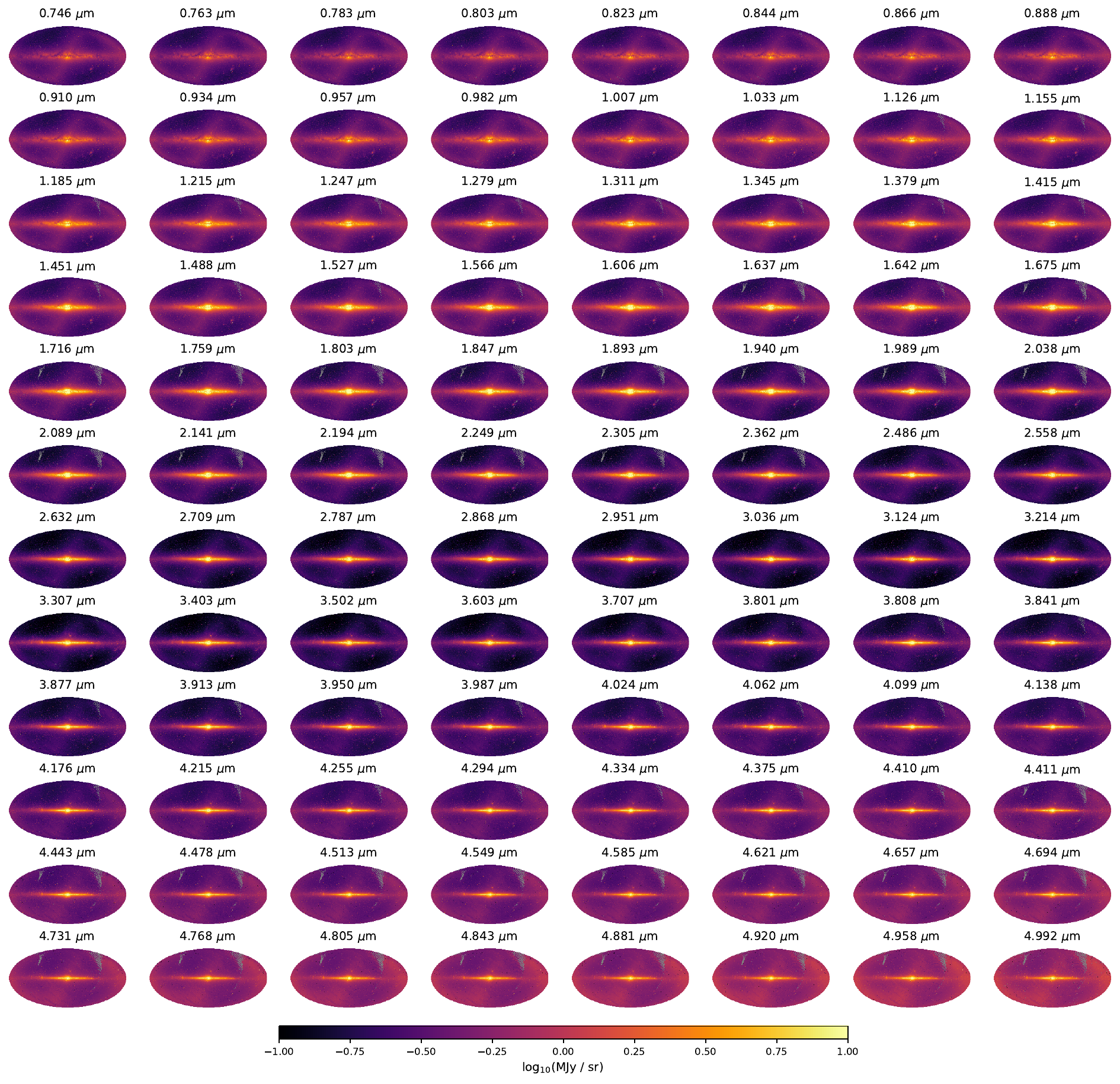}
    \caption{Full-sky maps in Mollweide projection in Galactic coordinates for \nChsFullSky\ spectral channels. We have removed the channels that directly border the DBS and the channels that are contaminated by helium airglow. No image-level filtering has been applied, so zodiacal light persists in the maps, most visibly as an S-curve corresponding to the ecliptic plane. The images have been smoothed to~$\fwhmFullSky$ in order to roughly match the $\resolFullSky$~pixels~($\nside = \nsideFullSky$). Gray pixels indicate missing coverage as of \fullskydate.}
    \label{fig:fullsky_datacube}
}
Whereas \figref{fig:datacube} focused on $\dimsCatsEye$ around the \catseye, the full-sky data cube represents the opposite extreme of \spherex's map-making capabilities. As for the \catseye, we have removed the two channels neighboring the~DBS, and we have additionally removed the four channels that are most affected by helium airglow~(near~$\wlHe~\um$). No image-level filtering has been applied, so the airglow is unmitigated. In \figref{fig:deproj_fullsky}, we feature the $\wlDeproj$-$\um$ map in a rendering that is larger and more convenient for examination.

On the largest scales, the full-sky maps are dominated by zodiacal light and the \milkyway. The zodiacal-light spectrum can be seen as the changing background glow; there is a transition from scattering~(at shorter wavelengths) to thermal emission~(at longer wavelengths) with a minimum near~$\wlZodiMin~\um$. The zodiacal light is brighter in the ecliptic plane, which can be seen as an S-curve stretching from the lower left to the upper right of each map. The \milkyway\ runs horizontally across each map. At shorter wavelengths, the Galactic plane is populated with dust absorption lanes. At longer wavelengths, the absorption weakens while the starlight spectrum is also changing. 

Bright stars contribute on smaller scales. These maps were made with $\nside=\nsideFullSky$, which corresponds to a pixel size of~$\resolFullSky$. To match this resolution, we smoothed the images to~$\fwhmFullSky$ before reprojection. Bright stars can still be seen in the maps, but they are substantially diluted. The Large and Small Magellanic Clouds are also significant contributors~(toward the bottom right of each map).

\subsection{Deprojection of Zodiacal Light \label{sec:fullskydeproj}}

We demonstrate deprojection~(\secref{sec:deproj}) of zodiacal light from the full-sky maps. Although the zodiacal light is a scientific target in its own right, we now consider it to be a foreground that we wish to minimize. In \figref{fig:deproj_fullsky}, we consider the $\wlDeproj$-$\um$~map, in which zodiacal light is most noticeable as a large-scale S-curve corresponding to the ecliptic plane. 
\figenv{
    \includegraphics[width=\textwidth]{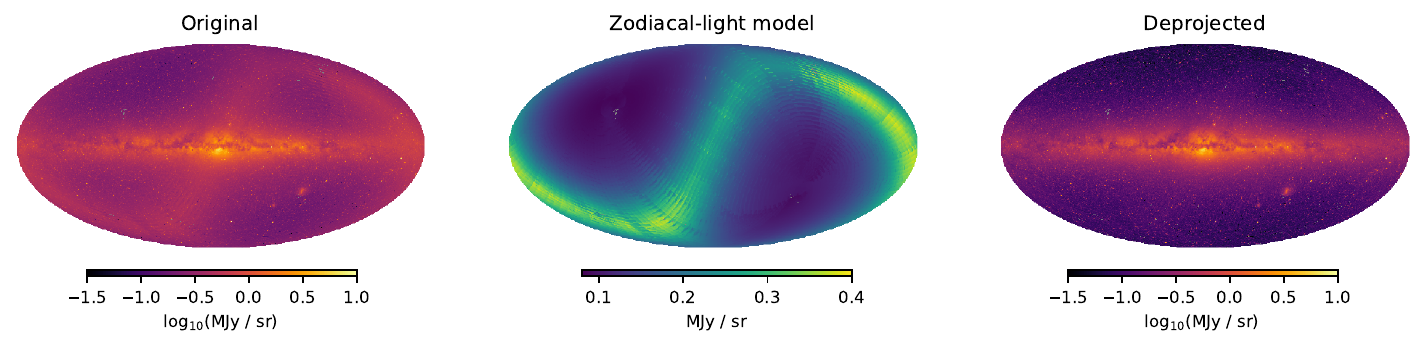}
    \caption{Deprojection of zodiacal light from a full-sky map. (\emph{Left})~The $\wlDeproj$-$\um$ full-sky map~(from the data cube of \figref{fig:fullsky_datacube}), which shows large-scale features associated with zodiacal light. (\emph{Middle})~The zodiacal-light model, which is based on simulations~(\secref{sec:skysim}). The jagged features are due to the time variability of zodiacal light as observed by \spherex~(cf.~ \figref{fig:sim_panel}). (\emph{Right})~The $\wlDeproj$-$\um$~map after the zodiacal-light model has been deprojected.  }
    \label{fig:deproj_fullsky}
}
Off of the ecliptic plane, zodiacal light contributes much of the background glow. On closer inspection, we find discrete discontinuities that are due to a combination of the scan strategy, the LVF~gradients and the time variability of the observed zodiacal light~(cf.~\figref{fig:sim_panel}).

In the example of \figref{fig:deproj_fullsky}, the deprojection template is a simulated map of zodiacal light as observed by \spherex~(\secref{sec:skysim}); the template is in the middle panel and is similar to the upper-left panel of \figref{fig:sim_panel}. Rather than subtracting the template directly, we fit for an overall amplitude before removal; this allows for some discrepancies between the real zodiacal light and the simulation. In order to accommodate non-zodiacal contributions to the real map, we fit simultaneously to a monopole, i.e., an overall offset, but we subtract only the best-fit zodiacal light. For the fit, we mask the Galaxy for $|b| < \latDeproj$, and we also mask bright sources. After the fitting procedure, the mask is lifted, and the deprojected map is presented in the right panel of \figref{fig:deproj_fullsky}, where the large-scale zodiacal-light features have been substantially suppressed. Importantly, the large-scale Galactic features have been retained.

\subsection{Full-sky Cross-correlations \label{sec:fullskycorr}}

As an illustration of the large-scale sensitivity and the convenience of the full-sky \healpix\ format, we perform a correlation analysis with tracers of interstellar dust. Our goal is to measure the spectrum of dust-correlated Galactic features in the near infrared. We mask bright sources in order to avoid disproportionate influence from a small number of sightlines, but a rigorous characterization of masking thresholds is beyond the scope of this work. Dimmer stars remain in the maps and contribute to the correlation. As a result, this example is different from measurements of the diffuse Galactic light~\citep[DGL, cf.][]{Brandt2012,Tsumura2013,Arai2015,Chellew2022}, although some methodological elements are similar. Diffuse components contribute but have not been isolated. A targeted study of DGL is left to future work. In this example, we measure the dust-correlated components of lightly-masked \spherex\ full-sky maps.

The first of our two dust tracers is the $\wlDust$-$\um$~map described by the \planck\ Collaboration in \cite{Planck2013XI}; this map is sensitive to thermal dust emission and was formed by combination of two different processings~\citep{Schlegel1998,Miville2005} of the observations of IRAS~\citep{Neugebauer1984} and DIRBE~\citep{Boggess1992}. One complication, however, is that far-infrared maps contain contributions from the cosmic infrared background~\citep[CIB,][]{Puget1996}. Without a dedicated removal process~\citep[e.g.,][]{Planck2016IntXLVIII,Chiang2023}, the CIB~component would correlate with the \spherex\ maps.

The second dust tracer is the map of \HI~column density~($\NHI$) that was presented in \cite{Lenz2017} and that is based on the dataset of \hiFourPi~\citep{HI4PI2016}. In the interstellar medium~(ISM), dust and \HI~gas are well-mixed~\citep[e.g.,][]{Boulanger1996}, and thermal dust emission is linearly related to 21-cm~\HI~emission for~$\NHI \lesssim \NHIlinear$. At higher densities, the \HI\ continues to correlate with dust emission but tends to produce an under-prediction, which may be due to an increasing fraction of~\Hmol. Because 21-cm~measurements can be separated into bins of line-of-sight velocity, extragalactic contributions can be mostly rejected by restricting to a maximum~\citep{Chiang2019}, which is~$90~\mathrm{km}/\mathrm{s}$ in this case.

To contrast with the investigation of \cite{Murgia2026}, which focuses on relatively bright regions near the Galactic plane, we consider a larger sky area at higher latitudes, where the ISM is relatively diffuse. We mask regions with $\NHI > \NHImask$, which retains \fskyHI~of the sky but allows for some nonlinearity in the relation between~\HI\ and dust emission. We also mask bright sources by sigma clipping to $\sigma=2$ based on the maps with deprojected zodiacal light. We note, however, that dimmer stars remain in the maps and partially correlate with dust anisotropies.

In \figref{fig:fullsky_dust_corr}, we present an example measurement of the dust-correlated components. 
\figenv{
     \gridline{
         \fig{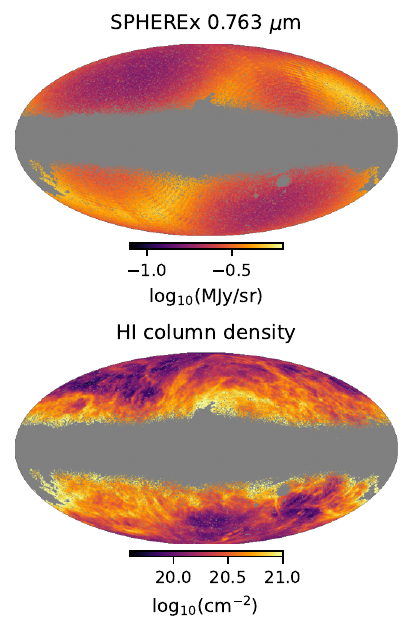}{0.30\textwidth}{}
        \fig{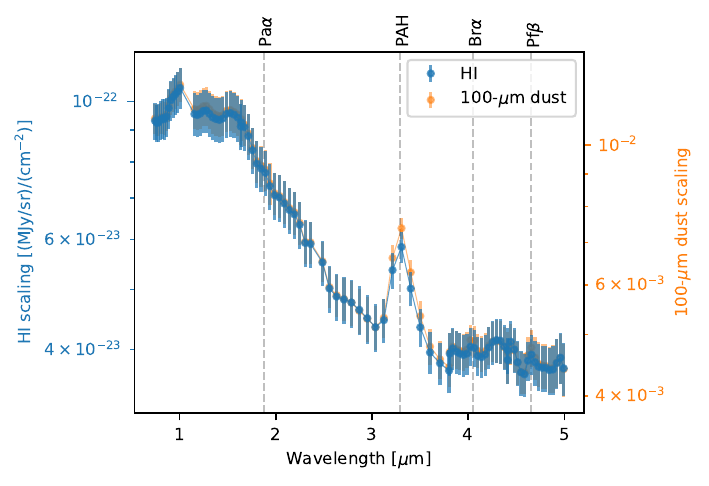}{0.68\textwidth}{}
    }
    \caption{Example spectrum of dust-correlated components at high latitudes. Two dust tracers were correlated with the full-sky data cube of \figref{fig:fullsky_datacube}. A mask was set by requiring $\NHI \leq \NHImask$ in the active areas. (\emph{Top left})~Example \spherex\ map with the mask applied. At high latitudes, the large-scale features are dominated by zodiacal light, but the correlation with dust tracers is able to extract signals that are relatively faint. (\emph{Bottom left})~The $\NHI$~map with the same mask; this map is used as one of the two dust tracers. (\emph{Right})~Spectra from correlating \spherex\ with the $\NHI$~map~(blue) and the $\wlDust$-$\um$~dust map~(orange). The spectra are reported as scaling factors, which convert the dust templates to the correlated components in the \spherex\ maps. The $\wlDust$-$\um$~map, which is not shown in the figure, has the same units~($\MJysr$) as the \spherex\ maps, so the scaling factors are dimensionless. Error bars represent variance estimated from dividing the sky into bins of Galactic longitude.}
    \label{fig:fullsky_dust_corr}
}
In the upper left, we provide an example \spherex\ map with our chosen mask; the remaining sky is dominated by zodiacal light. Below is the $\NHI$~map with the same mask. The $\wlDust$-$\um$~dust map is similar but has been omitted from the figure. For each \spherex\ wavelength map~(\figref{fig:fullsky_datacube}), we perform a simultaneous fit to zodiacal light~(as in \figref{fig:deproj_fullsky}), a monopole and one of the two dust tracers. The zodiacal-light simulation depends on wavelength, but the other templates do not. We form a spectrum from the best-fit amplitudes for dust. 

To estimate variance, we divide the sky into \nlonBins~bins of Galactic longitude. We measure the spectrum for each bin, and the standard deviations are used in \figref{fig:fullsky_dust_corr} to form error bars for the mean signal. Each longitude bin captures a range of variation associated with the multi-phase~ISM, which could be characterized with studies that are more localized. In this example, we estimate the average global signal.

In the right half of \figref{fig:fullsky_dust_corr}, we present the spectra obtained from both~\HI\ and $\wlDust$-$\um$~dust. The results are visually consistent for the two dust tracers. From short to long wavelengths, there is a steep decline, which may be a measure of the dust scattering efficiency. We find a strong $\wlPAH$-$\um$~PAH feature and modest enhancements associated with \Bra~($\wlBra~\um$) and \Pfb~($\wlPfb~\um$). Although we have marked the location of \Paa~($\wlPaa~\um$), the signal is weaker, perhaps because the spectral resolution is coarser in array~3~(\secref{sec:spectral_images}). Some small but abrupt shifts can be seen at the array boundaries, e.g., at $\sim 3.8~\um$, where array~4 meets array~5. These may be due to differences in sky coverage or atmospheric contamination, which remains unmitigated in this example.

The large-area cross-correlations are able to extract relatively faint dust signals. The wavelengths are highly correlated, and the longitude binning reveals substantial spatial variation, which may be an indication of interesting environmental influences. This investigation could be continued by studying how the spectrum depends on dust density, Galactic latitude, angular scale, dust tracer, survey coverage, masking,~etc. Our present purpose, however, is simply to provide an example of a measurement that is enabled by \spherex\ full-sky maps, so we defer further analysis to future work.

\subsection{Smoothing \label{sec:smoothing}}

Because the full-sky maps generally have coarser pixelizations than the \spherex\ detectors, we smooth the images before reprojecting to the map grid. Smoothing acts as an anti-aliasing filter and ensures that the map grid can properly sample the information that is reprojected to it. We smooth with a Gaussian kernel that is parameterized by its full width at half maximum~($\fwhm$). Typically, the~FWHM is set to at least twice the map pixel size. For the examples in \figs{\ref{fig:fullsky_datacube}, \ref{fig:deproj_fullsky} and~\ref{fig:fullsky_dust_corr}}, the~$\fwhm$ was set to~$\fwhmFullSky$ to accommodate a \healpix\ grid with $\nside=\nsideFullSky$, which corresponds to a map pixel size of~$\resolFullSky$. Although the following examples involve \healpix, it should be noted that smoothing can be applied for reprojection to any grid. Smoothing is most relevant for studies of large sky areas, for which fine-scale information can be neglected.

In \figref{fig:smooth}, we provide an example of image-level smoothing that is intended for a \healpix\ grid with $\nside=\nsideSmooth$~(pixel size of~$\resolSmooth$).
\figenv{
    \includegraphics[width=\textwidth]{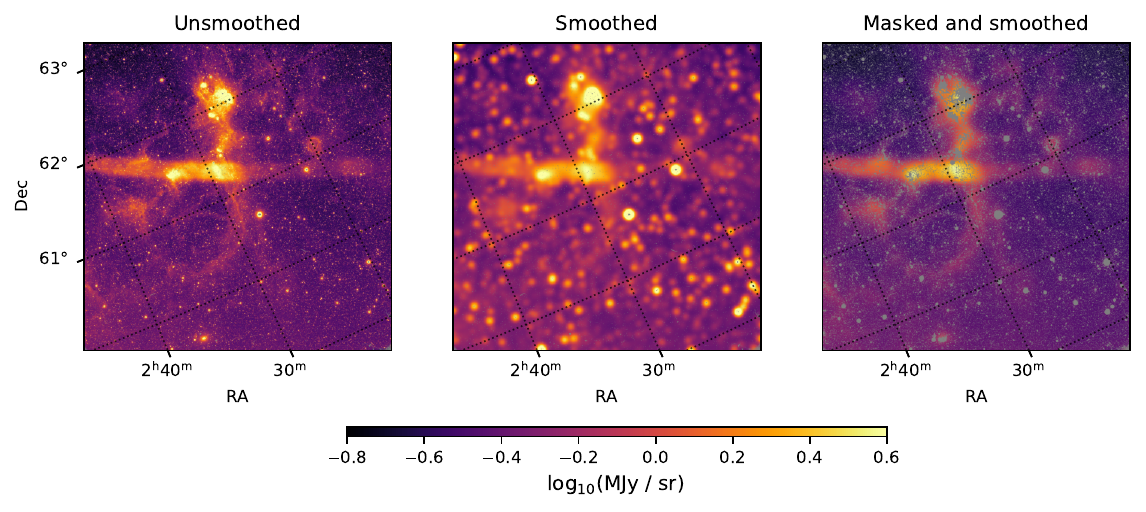}
    \caption{Examples of image-level smoothing and point-source removal. All of the images have been masked for pixel flags~(\secref{sec:pixel_masking}), which often affect the centers of bright stars. (\emph{Left})~An array-5 spectral image centered near the \heart~(identical to the bottom middle panel of \figref{fig:exp_panel}). (\emph{Middle})~The same image with smoothing to $\fwhm=\fwhmSmooth$. (\emph{Right})~The same image with source masking applied before smoothing; diffuse features are retained, and sources are suppressed.}
    \label{fig:smooth}
}
This pixelization is finer than for the previous examples, but it is still much coarser than \spherex's detector pixels~($\pixsize$).
On the left of \figref{fig:smooth}, we reproduce the array-5 spectral image from \figref{fig:exp_panel}. In the middle panel, the image has been smoothed with $\fwhm = \fwhmSmooth$. Fine features are now heavily suppressed, and bright stars occupy larger areas. The \Bra\ enhancement, which creates a sharp horizontal band in the unsmoothed image, has been softened, and this translates to a partial mitigation of LVF~gradients~(\secref{sec:specInterp}) in the mosaics.

In the upper left of \figref{fig:smoothcoadd}, we present a mosaic constructed from images that have been smoothed like the middle panel of \figref{fig:smooth}.
\figenv{
    \includegraphics[width=\textwidth]{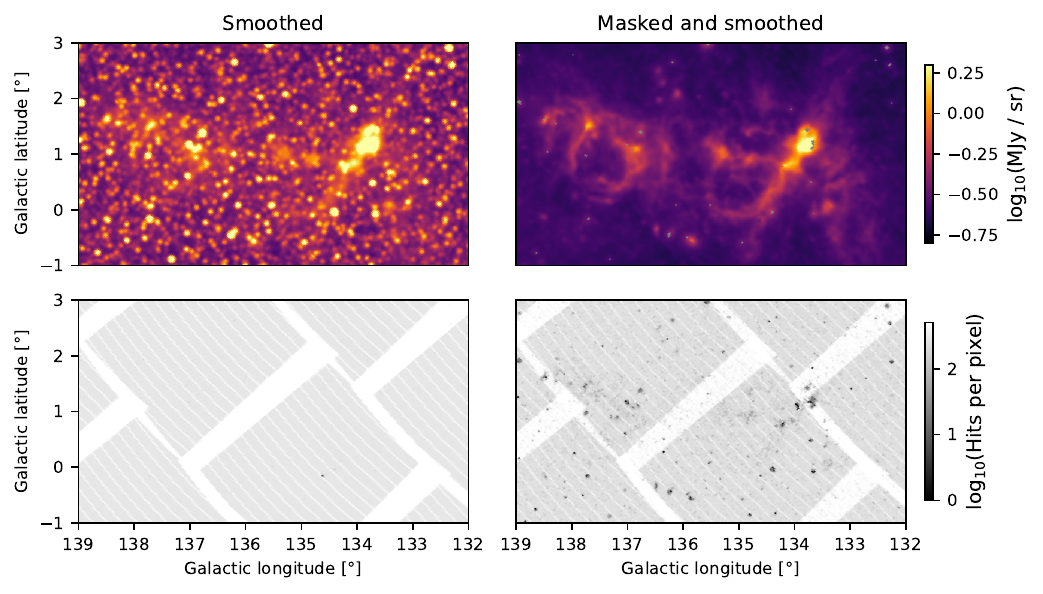}
    \caption{Example mosaics resulting from image-level smoothing. This is the same sky region as in the examples of \figref{fig:coadd_panel}, but the grid is now defined by \healpix\ with $\nside=\nsideSmooth$. (\emph{Top left})~With image-level smoothing~(\secref{sec:smoothing}, middle of \figref{fig:smooth}), the map is dominated by bright stars. (\emph{Bottom left})~The corresponding hits map, which shows that hundreds of image pixels~($\pixsize$) contribute to each mosaic pixel~($\resolSmooth$) in this example. (\emph{Top right})~With image-level source masking followed by smoothing~(\secref{sec:sourceremoval}, right of \figref{fig:smooth}), the map is dominated by structures that are more diffuse. (\emph{Bottom right})~The corresponding hits map, which shows diminishments at the locations of bright sources.}
    \label{fig:smoothcoadd}
}
The corresponding unsmoothed mosaic is in the upper right of \figref{fig:coadd_panel}. Unlike that previous example, we now reproject to a \healpix\ grid. The hits map is provided in the lower left of \figref{fig:smoothcoadd}; the scan patterns are similar to those of \figref{fig:coadd_panel}, but the number of hits per pixel is much larger due to the coarser pixelization~($\resolSmooth$ vs.~$\pixsizeCoadd$). As a result of the smoothing, the mosaic is dominated by bright stars, which may be desirable for studies of, e.g., integrated starlight~(ISL).

\subsection{Point-source Removal \label{sec:sourceremoval}}

For the study of diffuse structures, point sources may be masked before smoothing is applied. This technique is used in \cite{Murgia2026} for studies of PAH and \Bra\ structures in the Galactic plane. The operations are illustrated in the right panel of \figref{fig:smooth}, where point sources have been largely suppressed but diffuse structures remain. A source mask suppresses but does not eliminate the contribution of point sources, so the results depend on the specific masking choices~(\secref{sec:pixel_masking}).  Another approach, which is beyond the scope of this work, is to develop a PSF~model with which known sources can be deprojected from either the images or the mosaics.

In the upper right of \figref{fig:smoothcoadd}, we present a mosaic constructed from images that have been masked and smoothed like the right panel of \figref{fig:smooth}. The result is dominated by features that are diffuse and nebular, and bright sources have been heavily suppressed. In each image~(e.g., the right panel of \figref{fig:smooth}), the source mask renders many detector pixels inactive, but sky coverage can be maintained when reprojecting to a grid that is relatively coarse. For the examples of \figref{fig:smoothcoadd}, the grid resolution is~$\resolSmooth$, so approximately $280$~image pixels are associated with each mosaic pixel. In the lower right of \figref{fig:smoothcoadd}, we provide the hits map for the case of image-level source masking. When the hits count is exactly zero, the corresponding mosaic pixel is gray in the upper right panel. At the locations of bright sources, the hits are diminished but mostly non-zero. 

\section{Conclusion}

While this work provides a reference for the basic architecture of \spherex\ map making, improvements are expected and already under development. Foreground modeling is benefiting from \spherex\ data itself. Templates are under construction for the strongest atmospheric signals including airglow, aurorae and shuttle glow~\citep{Bock2026,airglow2026}. The zodiacal-light model~\citep{Crill2025} is being refined by comparison with \spherex\ observations. With better atmospheric and zodiacal-light models, image-level filtering~(\secref{sec:contourFilter}) can be more effective and targeted; zodiacal light could also be more efficiently deprojected from the mosaics~(\secref{sec:fullskydeproj}). Point sources are being modeled with increasing fidelity as a result of in-flight photometry and PSF~characterization~\citep{Akeson2025}; this will improve image-level source masking~(\secref{sec:pixel_masking}) and mosaic-level deprojection~(\secref{sec:deproj}). For the study of diffuse structures, it may also be desirable to perform point-source deprojection at the level of the images, i.e., before reprojecting to the mosaic grid; this is an approach that was mentioned in \secref{sec:sourceremoval} but deferred to future work.

Improved maps and methods will also enable improvements in the data products that are derived from those maps. With a reduction in foreground contamination, mosaic-level spectroscopy~(\secref{sec:spectroscopy}) can be more sensitive, and spectral features can be more reliably separated from continuum emission~(\secref{sec:continuum_subtraction}). As the data properties are better understood, specific signals can be extracted by fitting spectral templates to the maps; this is to be contrasted with the model-independent method of spectral interpolation, which was described in \secref{sec:specInterp} in order to suppress the LVF~gradients.

This article has been mostly restricted to the dataset acquired before \fullskydate, which represents approximately one quarter of \spherex's 2-year main mission. Importantly, this initial dataset represents a nearly complete survey of the full sky~(\secref{sec:scan} and \figref{fig:hits_panel}). At the time of writing, \spherex\ is revisiting most regions of the sky with different detector orientations, and subsequent full-sky surveys will introduce interleaved spectral sampling. With a combination of redundancy and systematic variation, \spherex\ will achieve greater raw sensitivity and enable the extraction of finer spectral features. Furthermore, continued observations allow for the recovery of sightlines that were affected by transients and other stochastic pathologies.

We have presented map-making methods that may be employed in a variety of investigations that involve data from \spherex. At the time of writing, a few map-related results have already been reported~\citep{Lisse2026,Hora2026,Murgia2026}. Many more investigations are planned by the \spherex\ \st, and some are currently in progress. Future publications may be accompanied by the release of maps and spectra. A public mosaic tool will be made available by \irsa~\citep{Akeson2025}, and the broader community is likely to find additional science goals that can be advanced by the types of maps described in this work.

\begin{acknowledgments}
We acknowledge support from the SPHEREx project under a contract from the NASA/Goddard Space Flight Center to the California Institute of Technology.
Part of the research described in this paper was carried out at the Jet Propulsion Laboratory, California Institute of Technology, under a contract with the National Aeronautics and Space Administration~(80NM0018D0004).

The authors acknowledge the Texas Advanced Computing Center~(TACC)\footnote{\url{http://www.tacc.utexas.edu}} at The University of Texas at Austin for providing computational resources that have contributed to the research results reported within this paper.

We thank Brandon Hensley for advice and helpful conversations.
\end{acknowledgments}

\bibliography{MM}{}
\bibliographystyle{aasjournalv7}

\end{document}